\documentclass[twocolumn]{aastex631}
\accepted{October 17, 2024}

\begin{document}

\title{Filament Accretion and Fragmentation in the Perseus Molecular Cloud}

\author[0000-0003-4242-973X]{Michael Chun-Yuan Chen}
\affiliation{Herzberg Astronomy and Astrophysics, National Research Council of Canada, 5071 West Saanich Road, Victoria, BC, V9E 2E7, Canada}

\author[0000-0002-9289-2450]{James Di Francesco}
\affiliation{Herzberg Astronomy and Astrophysics, National Research Council of Canada, 5071 West Saanich Road, Victoria, BC, V9E 2E7, Canada}
\affiliation{Department of Physics and Astronomy, University of Victoria, 3800 Finnerty Road, Elliott Building, Room 101, Victoria, BC , V8P 5C2, Canada}

\author[0000-0001-7594-8128]{Rachel K. Friesen}
\affiliation{Department of Astronomy, and Astrophysics, University of Toronto, 50 St. George Street (Room 101), Toronto, ON, M5S 3H4, Canada}

\author[0000-0002-3972-1978]{Jaime E. Pineda}
\affiliation{Max-Planck-Institut für extraterrestrische Physik, Giessenbachstrasse 1, D-85748 Garching, Germany}

\author[0000-0003-1481-7911]{Paola Caselli}
\affiliation{Max-Planck-Institut für extraterrestrische Physik, Giessenbachstrasse 1, D-85748 Garching, Germany}

\author[0000-0001-6431-9633]{Adam Ginsburg}
\affiliation{Department of Astronomy, University of Florida, P.O. Box 112055, Gainesville, FL, USA}

\author[0000-0002-5779-8549]{Helen Kirk}
\affiliation{Herzberg Astronomy and Astrophysics, National Research Council of Canada, 5071 West Saanich Road, Victoria, BC, V9E 2E7, Canada}

\author[0000-0001-6004-875X]{Anna Punanova}
\affiliation{Ural Federal University, 620002, 19 Mira street, Yekaterinburg, Russia}

\collaboration{20}{GAS Collaboration}
\noaffiliation



\begin{abstract}

Observations suggest that filaments in molecular clouds can grow by mass accretion while forming cores via fragmentation. Here we present one of the first large sample studies of filament accretion using velocity gradient measurements of star-forming filaments on the $\sim 0.05$ pc scale with NH$_3$ observations of the Perseus Molecular Cloud, primarily obtained as a part of the GBT Ammonia Survey (GAS). In this study, we find significant correlations between velocity gradient, velocity dispersion, mass per unit length, and the number of cores per unit length of the Perseus filaments. Our results suggest a scenario in which filaments not only grow through mass accretion but also form new cores continuously in the process well into the thermally supercritical regime. Such behavior is contrary to that expected from isolated filament models but consistent with how filaments form within a more realistic cloud environment, suggesting that the cloud environment plays a crucial role in shaping core formation and evolution in filaments. Furthermore, even though velocity gradients within filaments are not oriented randomly, we find no correlation between velocity gradient orientation and the filament properties we analyzed. This result suggests that gravity is unlikely the dominant mechanism imposing order on the $\sim 0.05$ pc scale for dense star-forming gas.

\end{abstract}



\section{Introduction} \label{sec:intro}

Star formation is intimately linked to filaments in molecular clouds \citep{Andre2014, Pineda2023, Hacar2023}. Observationally, filaments are both ubiquitous in molecular clouds (e.g., \citealt{Andre2010}) and host a majority of the star-forming cores \citep{Konyves2015}. Theoretically, filaments can be naturally produced by supersonic turbulence in molecular clouds (e.g., \citealt{Porter1994}; \citealt{Vazquez-Semadeni1994}) and can undergo collapse to form stars under self-gravity (e.g., \citealt{Ostriker1999}; \citealt{Ballesteros-Paredes1999}; \citealt{MacLow2004}). The details on how molecular clouds assemble filaments and subsequently produce stars, however, are not well-constrained currently. 

The formation of dense cores, the direct progenitor of protostars \citep{DiFrancesco2007}, can often occur through filament fragmentation \citep{Andre2014}. Under this scenario, filaments with masses per unit length ($M_{\mathrm{lin}}$) above the thermally critical value of $M_{\mathrm{lin, crit}} \sim 16$ M$_\odot$ pc$^{-1}$ (\citealt{Ostriker1964}) at an isothermal temperature of 10 K are expected to collapse or fragment when modeled as cylinders near hydrostatic equilibrium. Consequently, the perturbations that grow the fastest during the thermally subcritical phase are expected to determine the fragmentation length scale in both quasi-equilibrium (e.g., \citealt{Inutsuka1992}) and non-equilibrium (e.g., \citealt{Clarke2016}) cylindrical models. 

More realistic simulations that include the cloud environment can produce filaments and cores simultaneously (e.g., \citealt{ChenCheYu2015}). The core properties found in these models are often governed predominantly by cloud characteristics such as turbulent pressure and magnetic field strength. In the magnetized cases, the mass flow towards cores along the filaments, typically expected from the fragmentation models, can be significantly suppressed by the magnetic fields. Accretion towards cores and filaments in these models is thus expected to flow predominantly along the magnetic field lines, which are not necessarily perpendicular to the filaments, until cores become sufficiently massive at a later stage (e.g., \citealt{ChenCheYu2014}).

The ability to probe accretion flows with respect to the local orientation of the filaments is thus crucial to constraining core and filament formation models. Such constraints on the accretion flow behavior can be further complemented by measurements of velocity dispersion ($\sigma_v$) to probe accretion-driven turbulence \citep[e.g.,][]{Klessen2010, Heigl2020}, $M_{\mathrm{lin}}$ to infer filament growth \citep[e.g.,][]{Chira2018, ChenCheYu2020, Feng2024}, and average core spacing to test fragmentation theories (e.g., \citealt{Inutsuka1992}; \citealt{Clarke2016}). Indeed, the large-scale velocity field that runs perpendicular to the Taurus B211/3 filament and parallel to the magnetic field lines indicates that this thermally supercritical ($M_{\mathrm{lin}} \sim 54$ M$_\odot$ pc$^{-1}$) filament is accreting from its surroundings \citep{Palmeirim2013}. Moreover, larger sample studies of filaments in the IC 5146, Aquila, and Polaris clouds have shown a correlation between $M_{\mathrm{lin}}$ and $\sigma_v$ in thermally supercritical filaments, which further suggests the prevalence of filament growth through accretion \citep{Arzoumanian2013}.

While many velocity gradient ($\nabla v_{\mathrm{LSR}}$) analyses of filaments have been made along (e.g., \citealt{KirkHelen2013}; \citealt{Friesen2013}) and across (e.g., \citealt{Fernandez-Lopez2014}; \citealt{Dhabal2018}) filaments on the global scale, relatively few studies have measured $\nabla v_{\mathrm{LSR}}$ at the beam resolution (e.g., \citealt{Hacar2018}). Measurements of small-scale $\nabla v_{\mathrm{LSR}}$ orientations on the plane of the sky, which have been shown to be fairly complex, are fewer still (e.g., \citealt{ChenM2020}, \citeyear{ChenMCY2022, ChenMike2024}). Here, we present one of the first systematic, large-sample studies of the $\nabla v_{\mathrm{LSR}}$ fields measured on smaller scales ($\sim 0.05$ pc), accompanied by measurements of $\sigma_v$, $M_{\mathrm{lin}}$, number of cores per unit length, and number of embedded protostars per unit length using observations of the Perseus Molecular Cloud.

At a distance of $\sim 300$ pc \citep{Zucker2018}, the Perseus Molecular Cloud is one of the most well-studied nearby star-forming clouds. The cloud's dense core population, for example, has been systematically studied with both dust continuum (e.g., \citealt{Enoch2006}; \citealt{KirkHelen2006}; \citealt{Sadavoy2010}) and molecular observations (e.g., \citealt{KirkHelen2007}; \citealt{Rosolowsky2008}). Numerous surveys have also revealed the full cloud structure of Perseus in dust continuum (e.g., \citealt{Pezzuto2021}) and molecular emission observations in $^{12}$CO and $^{13}$CO (e.g., \citealt{Ridge2006}). These full-cloud surveys are later followed up by high-resolution mapping of individual star-forming clumps with tracers such as HCO$^+$ and N$_2$H$^+$ (e.g., \citealt{Walsh2007}; \citealt{Storm2014}), as well as NH$_3$ (e.g., \citealt{Pineda2011}, \citeyear{Pineda2015}; \citealt{Dhabal2019}).

We lay out our paper as follows: we present the observational details of our NH$_3$ data in Section \ref{sec:data} followed by methods on our (up to) two-component spectral fitting as well as our filament identification and analyses in Section \ref{sec:method}. We present and discuss our results in Section \ref{sec:discussion} and present a summary of our findings, accompanied by concluding remarks, in Section \ref{sec:summary}.

%============================================================
\section{Data} \label{sec:data}

\begin{figure*}
\centering
\includegraphics[width=1.0\textwidth]{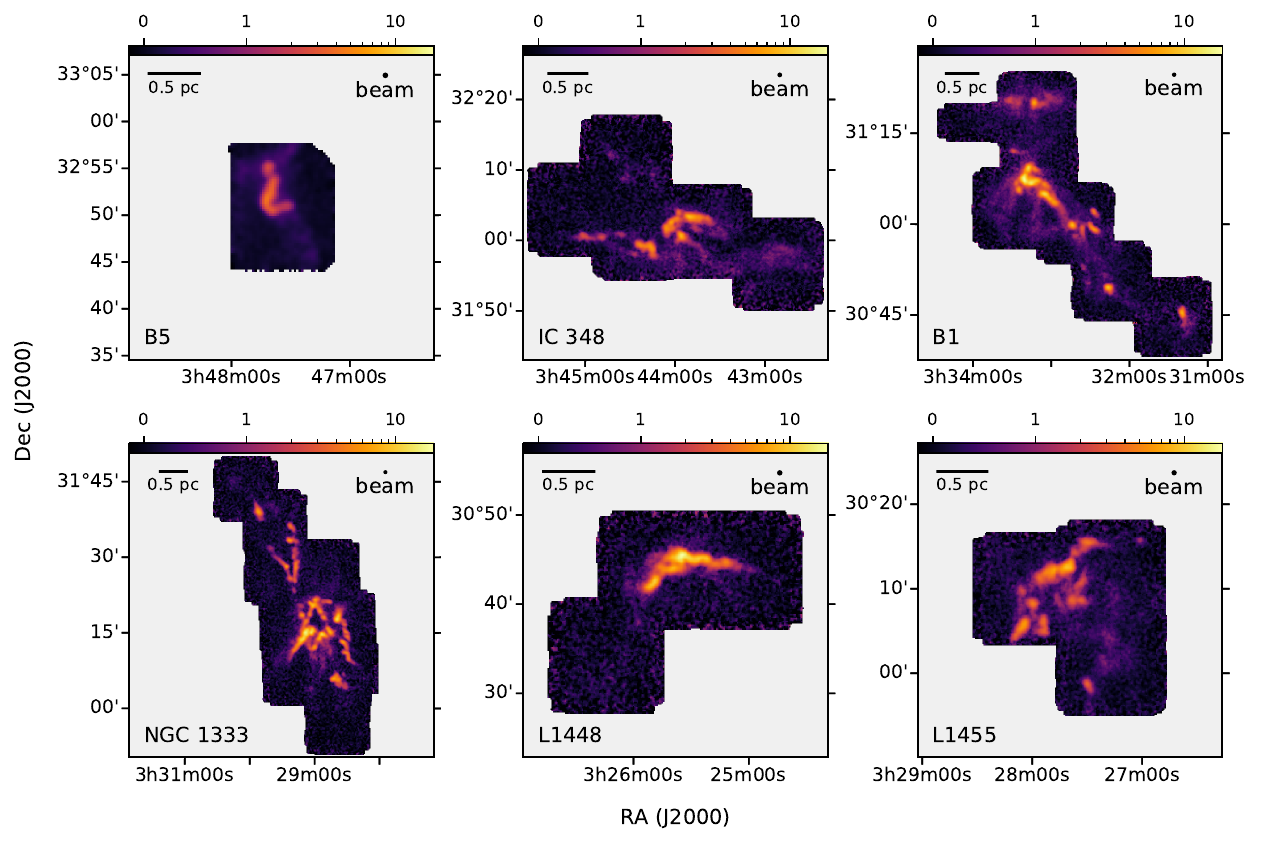}
\caption{The NH$_3$ (1,1) integrated intensity maps of the six Perseus clumps in units of K km s$^{-1}$. The scale bar and beam size of each map are shown in the top left and top right corners of their panels, respectively. \label{fig:NH3_mom0}}
\end{figure*}

We obtained our NH$_3$ data primarily as a part of the Green Bank Ammonia Survey (GAS; \citealt{Friesen2017}), which mapped star-forming regions in the Gould Belt molecular clouds visible to the northern hemisphere with A$_{\mathrm{V}} > 7$ mag. The GAS team observed the NH$_3$ (1,1) and (2,2) inversion lines in these regions with the Robert C. Byrd Green Bank Telescope (GBT) using its 7-beam K-Band Focal Plane Array (KFPA) and its VErsatile GBT Astronomical Spectrometer (VEGAS) backend. The full width at half maximum (FWHM) spectral resolution of these observations is 5.7 kHz, corresponding to $\sim 0.07$ km s$^{-1}$ at 23.7 GHz. The FWHM beam of the GBT at this frequency is $32''$ and corresponds to a physical resolution of $\sim 0.05$ pc at a 300 pc distance. The GBT's main beam efficiency is 0.81 at this frequency.

The GAS observations were conducted over $10' \times 10'$ on-sky tiles using the On-The-Fly (OTF) technique, which scans the sky along the right ascension (R.A.) direction with Nyquist-sampled spacing between each row. We reduced and imaged these observations to main-beam brightness temperature ($T_\mathrm{MB}$) units with the GBT KFPA data reduction pipeline \citep{Masters2011} using a recipe by \cite{Mangum2007}. Further details of the GAS data products are available in the Data Release 1 \citep[DR1;][]{Friesen2017} and the Data Release 2 (DR2; Pineda et al. submitted) papers.

Additionally, we include the GBT NH$_3$ observations of the Perseus B5 clump by \cite{Pineda2010} in our study, which is not covered by GAS. The B5 data were obtained similarly using the \textit{K}-band receiver via the OTF technique and imaged with the recipe by \cite{Mangum2007}. The resulting data product has a 3.05 kHz resolution, equivalent to 0.04 km s$^{-1}$ at 23.7 GHz. Further details of the data can be found in the paper by \cite{Pineda2010}.

Figure \ref{fig:NH3_mom0} shows the NH$_3$ (1,1) integrated intensity maps of the six Perseus clumps included in this study, i.e., B5, IC 348, B1, NGC 1333, L1448, and L1455. The median channel rms of these data are 0.05 K, 0.12 K, 0.13, 0.11, 0.14, and 0.11, respectively. The overall channel rms are relatively uniform throughout the maps, except near the edges where sparse and variable receiver coverage produced significantly higher noise levels. The B5, NGC 1333, and the rest of the Perseus data are presented by \cite{Pineda2010}, \cite{Friesen2017}, and Pineda et al. (submitted), respectively.

%============================================================
\section{Methods \& Results} \label{sec:method}

%=============================
\subsection{Spectral Fitting} \label{subsec:spec_fit}

\begin{figure*}[t!]
\centering
\includegraphics[width=0.945\textwidth, trim={0 0.5mm 0 0}]{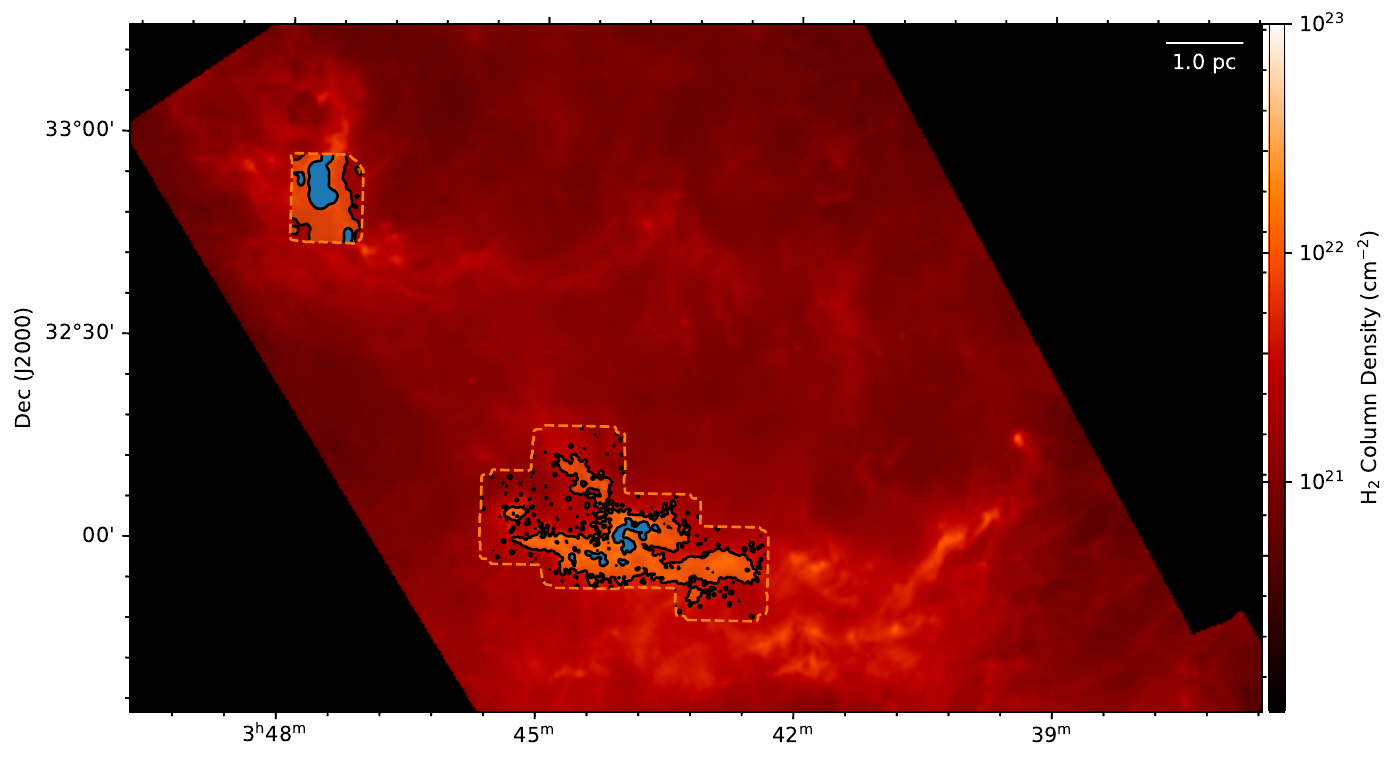}
\includegraphics[width=0.945\textwidth, trim={0 0 0 3mm}]{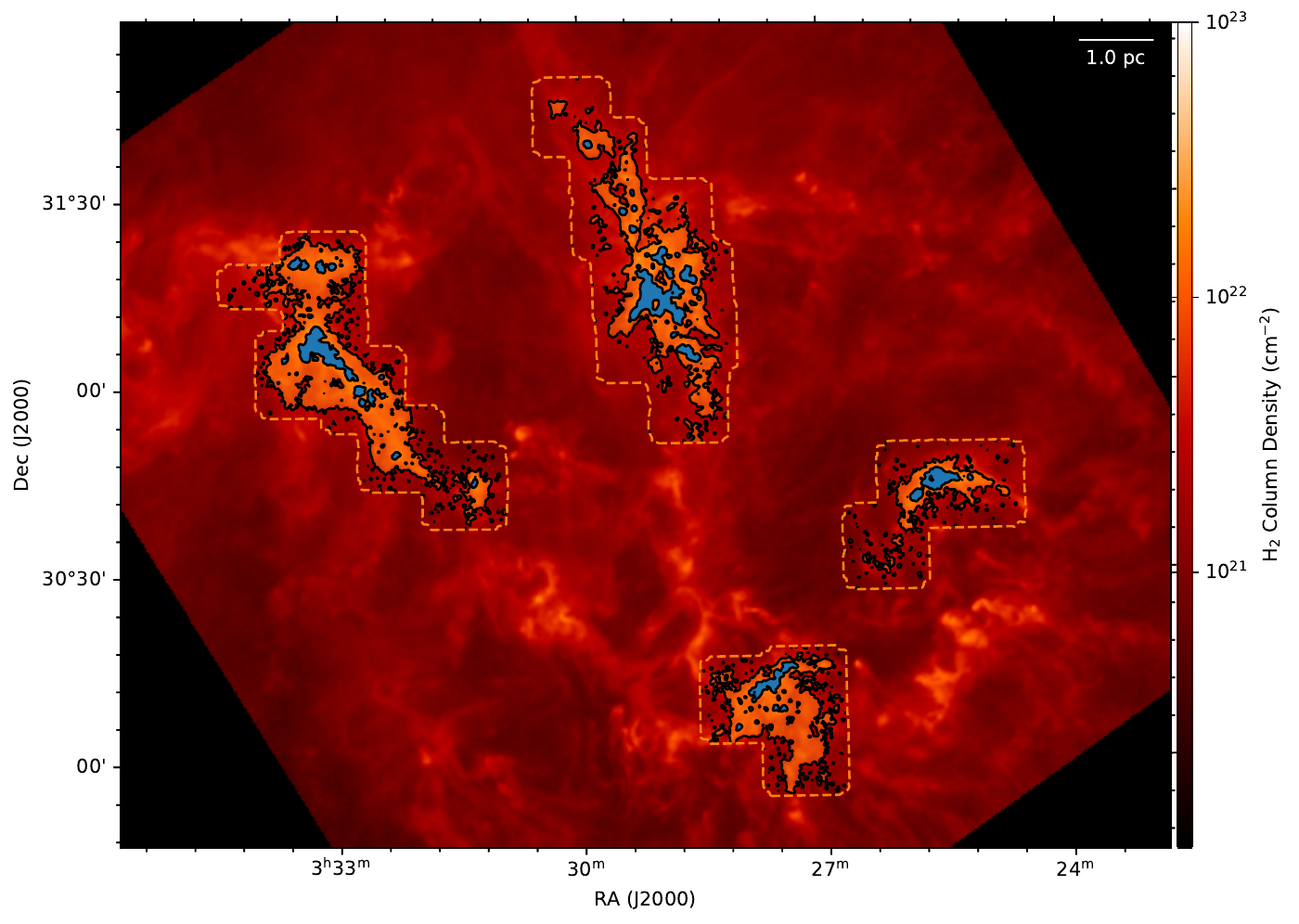}
\caption{The {\em Herschel}-derived H$_2$ column density map of Perseus East (top) and West (bottom), overlaid with filled contours showing where the GBT-observed NH$_3$ spectra are best-fitted by one-component and two-component models in orange and blue, respectively. The footprints of the GAS observations are shown in orange dashed contours. The scale bars are shown on the top right corners of each panel. \label{fig:Per_Map}}%\label{fig:PerWest_Map}
\end{figure*}

We fitted the NH$_3$ (1,1) spectral lines with either one or two-component models using the \texttt{MUFASA} software developed by \cite{ChenM2020}. \texttt{MUFASA} is an automated spectral fitter that utilizes the \texttt{PySpecKit} package (\citealt{Ginsburg2011}; \citealt{Ginsburg2022}) to perform least-squares spectral fitting using the Levenberg–Marquardt method (\citealt{Levenberg1944}; \citealt{marquardt1963}). \texttt{MUFASA} fits spectra on a per-pixel basis over two iterations. In the first iteration, \texttt{MUFASA} spatially smooths the data to twice its native beamsize and then fits the smoothed data using initial guesses derived from the corresponding moment maps. The spatial smoothing provides the fitting routine with a sensitivity boost and an increased spatial awareness over fits at the native resolution. In the subsequent iteration, the fitted parameters obtained from the first iteration are then used as initial guesses for the final, native-resolution fit. Further details of how \texttt{MUFASA} operates, including how initial guesses are derived from moment maps, can be found in \cite{ChenM2020}. Since the NH$_3$ (2,2) lines were detected over a much smaller area of the sky than the (1,1) lines, we do not include the NH$_3$ (2,2) lines in this study.

\begin{figure*}[t!]
\centering
\includegraphics[width=1.0\textwidth]{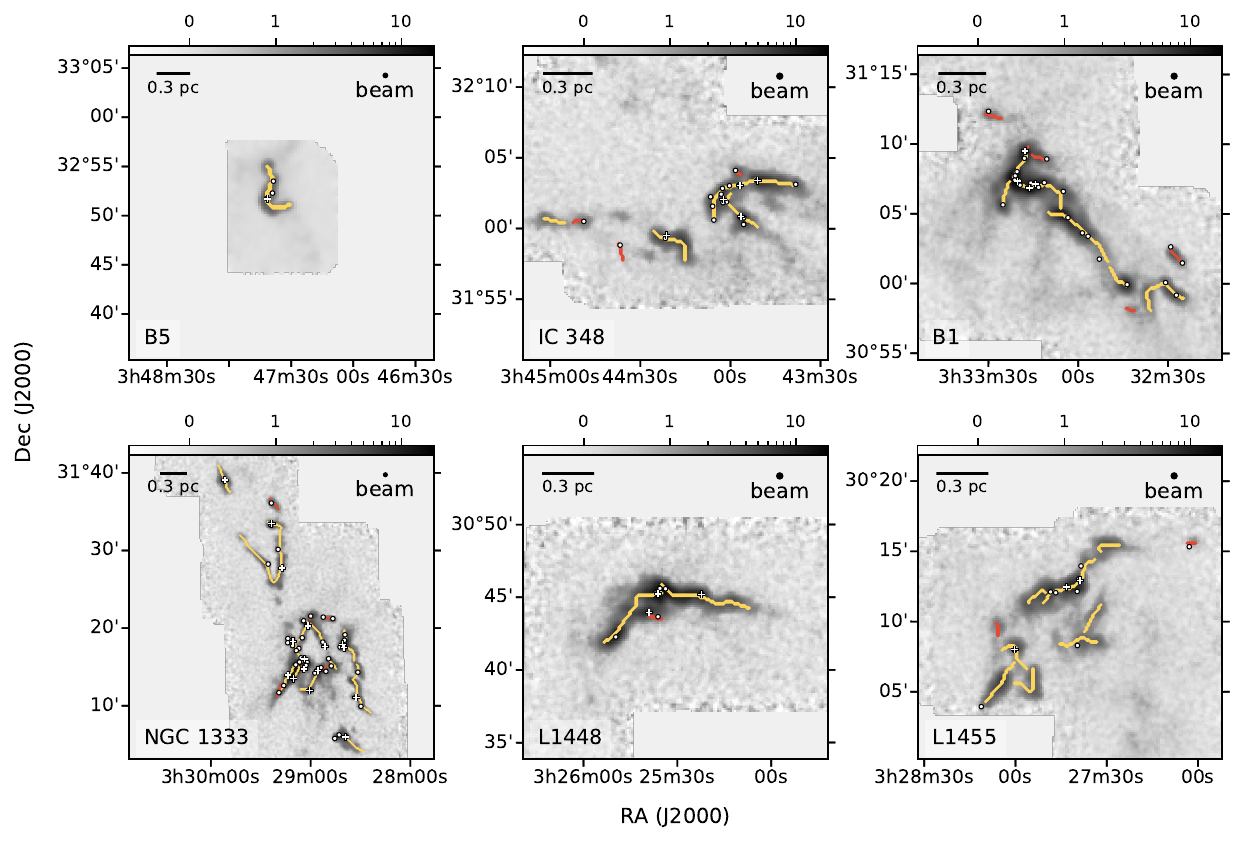}
\caption{The filament spines identified from the fits-derived NH$_3$ spectral models in PPV space, overlaid on top of the integrated intensity maps of the NH$_3$ (1,1) observations, which are in units of K km s$^{-1}$. Spines with lengths $\geq 10.5$ pixels (3 beamwidths) are shown in yellow, while the rest are shown in red. Positions of the cores identified by Pattle et al. (submitted) and Class 0/I YSOs identified by \cite{Dunham2015} are marked by circles and pluses, respectively, selected only for those located on our filaments. The remaining labels are the same as those in Figure \ref{fig:NH3_mom0}. \label{fig:Spine_Maps}}
\end{figure*}

Following \cite{ChenM2020}, we modeled each spectral component of the NH$_3$ (1,1) emission as a homogeneous slab of gas parameterized by its velocity centroid ($v_{\mathrm{LSR}}$), velocity dispersion ($\sigma_v$), excitation temperature ($T_{\mathrm{ex}}$), and optical depth ($\tau_0$). We assemble the spectral model by radiatively transferring the cosmic microwave background through each slab along a line of sight toward the observer. All 18 hyperfine structures of the NH$_3$ (1,1) emission are accounted for in our spectral model.

\texttt{MUFASA} selects the best-fit model from the two-component, one-component, and noise model fits using the corrected Akaike Information Criterion (AICc, \citealt{Akaike1974}; \citealt{Sugiura1978}), which accounts for a finite sample size of the data with a second-order correction. Specifically, \texttt{MUFASA} selects model \textit{b} over model \textit{a} when their AICc-determined log-likelihood, i.e.,
\begin{equation}\label{eq:lnk}
\ln{K_a^b} = - \left ( \textup{AICc}_b - \textup{AICc}_a \right )/2,
\end{equation}
is greater than a statistically robust value of 5 \citep{Burnham2004}. The AICc values are calculated as 
\begin{equation}\label{eq:AICc}
\mathrm{AICc} = n\ln{(\mathrm{RSS}/n) + 2p + \frac{2p(p+1)}{n-p-1}},
\end{equation}
where $n$ is the sample size, $p$ is the number of parameters, and $\mathrm{RSS}$ is the residual sum of squares. We note that \texttt{MUFASA}'s AICc-based model selection performed similarly to the full Bayesian approach over the same data (see \citealt{Sokolov2020}; \citealt{ChenM2020}), with the AICc-based method not having to search the likelihood space of the fits as exhaustively as in the Bayesian approach. Similar to the Bayesian approach, however, the AICc method is also capable of recovering fainter signals typically excluded by a method with a strict SNR cutoff (see \citealt{ChenM2020}). 

We note further that the form of AICc presented in Equation \ref{eq:AICc} has been updated from that presented by \citealt{ChenM2020}, which assumes the chi-squared-based likelihood estimator based on Wilks' theorem \citep{Wilks1938}. The updated form adopts the more commonly used least-squares likelihood estimator (e.g., \citealt{Burnham2004}) and has been implemented by \texttt{MUFASA} starting with version \texttt{v1.2.0}. The differences between the two implementations are marginal for our data, affecting only a small fraction of the pixels where $\ln{K_a^b} \sim 5$. For these marginal cases, the updated form has performed better on model selection.

Figure \ref{fig:Per_Map} shows the results of our \texttt{MUFASA} fits indicating where a one-component or two-component NH$_3$ (1,1) spectra best fit the data as determined by the AICc. The orange dashed lines show the extent of the GAS observations, overlaid on top of the N(H$_2$) (i.e., {\em Herschel}-derived H$_2$ column density) maps we used for our mass analyses. These $36''$ resolution N(H$_2$) data were taken from the {\em Herschel} Gould Belt Survey Archive and were first presented by \cite{Pezzuto2021}. The parameter maps derived from the one- and two-component fits, along with their associated AICc maps, are available at DOI 10.11570/24.0092. We have also included the $v_{\mathrm{LSR}}$ and $\sigma_{\mathrm{v}}$ maps of the six Perseus clumps in Appendix \ref{appendix:kin_maps}, with the two components sorted by their relative $\sigma_{\mathrm{v}}$ values in each pixel (i.e., along each line of sight).

%=============================
\subsection{Filament Identification} \label{subsec:fil_id}

We identify filaments from our fits-derived modeled emission in the position-position-velocity (PPV) space using the \texttt{CRISPy} software \citep{ChenM2020}. \texttt{CRISPy} identifies density ridges in multi-dimensional space using the Subspace Constrained Mean Shift algorithm (\citealt{Ozertem2011}; \citealt{ChenYC2014arXiv}; \citealt{ChenYC2015MNRAS}). The reconstructed emission cubes have hyperfine structures removed to ensure these cubes contain only real kinematic structures for \texttt{CRISPy} to identify. We accomplished this reconstruction by treating each already-fitted velocity slab's optical depth profile as a single Gaussian, with its peak optical depth re-normalized to one-tenth of the $\tau_0$ value derived from the full hyperfine structure fits to emulate the optical depths of the original satellite lines, which are optically thin in most cases. To optimize the spectral resolution for filament identification, we further replace the fits-derived $\sigma_v$ value of the model with a fixed value of 0.09 km s$^{-1}$, the minimally Nyquist-sampled value of the GAS data, in the reconstruction. Further details of such a reconstruction can be found in \cite{ChenM2020}.

The emission ridges \texttt{CRISPy} identified in continuous PPV space are then re-gridded back to the data's native grids. The branches on these ridges are subsequently pruned to produce branchless spines that trace the longest path in the ridges. Figure \ref{fig:Spine_Maps} shows \texttt{CRISPy}-identified spines overlaid on top of the NH$_3$ (1,1) integrated intensity maps. Spines with sky-projected lengths $\geq 10.5$ pixels and $< 10.5$ pixels are shown in yellow and red, respectively. The 10.5-pixel threshold corresponds to 3 beam widths, which we adopted as the minimum aspect ratio required to define a filament, as opposed to an elongated core, for our final analyses. In total, we identified 36 filaments in Perseus. We note that our results for NGC 1333 differ slightly from that of \cite{ChenM2020} due to recent improvements in the ridge-gridding method that reduce aliasing. 

We use the \texttt{CRISPy}-identified spines, both long and short, to sort our modeled spectral slabs into velocity-coherent structures, i.e., structures with $v_{\mathrm{LSR}}$ that are spatially continuous. We include short spines in this sorting process because they also trace kinematically coherent structures like their longer counterparts. We assign each velocity slab (i.e., component) at a given pixel a membership in a velocity-coherent structure based on its kinematic similarities to the filament spines and its neighboring slabs using the methods by \cite{ChenM2020}. To ensure we do not erroneously include ambient gas unrelated to a filament as a part of the membership, we spatially limit our filament's maximum extent to 5 pixels in radius from the spine, which corresponds to $\sim 0.1$ pc in Perseus. We note that this value is about twice the typical filament half-width found by {\em Herschel} studies, which have full-width-at-half-maximum (FWHM) values of $\sim 0.1$ pc (e.g., \citealt{Arzoumanian2011}; \citealt{Arzoumanian2019}), and about seven times that of the FWHM measured for Perseus B5 sub-filaments with NH$_3$ at higher resolutions ($\sim 0.03$ pc; \citealt{Pineda2011}; \citealt{Schmiedeke2021}; \citealt{ChenMCY2022}). Further details on how we assign memberships to velocity-coherent structures are presented by \cite{ChenM2020}.

\begin{figure}[t]
\centering
\includegraphics[width=1.0\columnwidth]{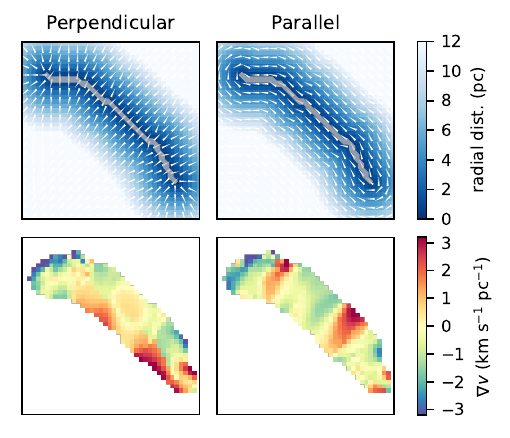}
\caption{The top left and right panels show example vector fields that are perpendicular or parallel to a filament spine, respectively, taken from a B1 filament. The filament spine and the distance between each pixel and the spine, from which the vector fields are derived, are shown in the background. The bottom left and right panels show the $\nabla v_{\mathrm{LSR}, \perp}$ and $\nabla v_{\mathrm{LSR}, \parallel}$ maps of the same filament, respectively. \label{fig:div_field_demo}}
\end{figure}

The velocity gradients ($\nabla v_{\mathrm{LSR}}$) within each velocity-coherent filament are calculated by fitting a plane to the fits-derived $v_{\mathrm{LSR}}$ values assigned to each filament within a 6-pixel diameter ($\sim 2$ beam widths) of a given position. We design such a measure to ensure the calculated $\nabla v_{\mathrm{LSR}}$ values are well-resolved by the beam. For a similar reason, we exclude $\nabla v_{\mathrm{LSR}}$ calculations at positions where more than one-third of the pixels are missing from within their 6-pixel diameter. Since we only calculate $\nabla v_{\mathrm{LSR}}$ within a velocity-coherent filament, our $\nabla v_{\mathrm{LSR}}$ are robust against calculating gradients across unrelated, overlapping objects on the plane of the sky.

Following \cite{ChenM2020}, we further decompose the $\nabla v_{\mathrm{LSR}}$ of each filament into components that are perpendicular and parallel to the spine. This decomposition is accomplished by taking dot products between the $\nabla v_{\mathrm{LSR}}$ and the vector fields that point away or along the filaments' spines. The former field is constructed by taking the gradient of pixel distances from the spine, while the latter field is constructed by rotating the former by $90 ^{\circ}$, as demonstrated in the top panels of Figure \ref{fig:div_field_demo}. The bottom panels of Figure \ref{fig:div_field_demo} show the resulting components, i.e., $\nabla v_{\mathrm{LSR}, \perp}$ and $\nabla v_{\mathrm{LSR}, \parallel}$, taken from a filament in the B1 clump as an example. We include $\nabla v_{\mathrm{LSR}, \perp}$ and $\nabla v_{\mathrm{LSR}, \parallel}$ maps of our Perseus filaments in Appendix \ref{appendix:vgrad_maps}.

To estimate the total mass of each filament, we spatially integrate {\em Herschel}-derived column densities over each velocity-coherent filament's on-sky footprint. These column density maps were first interpolated and regridded onto the respective NH$_3$ data's grid before the integration. To estimate the physical pixel sizes in each of the Perseus star-forming clumps, we assume the B5, IC 348, B1, NGC 1333, L1448, and L1455 clumps are at distances of 302 pc, 295 pc, 301 pc, 299 pc, 288 pc, and 279 pc away, respectively. We adopt these distances directly from the measurements of \cite{Zucker2018}, with the exception of L1455, which was not explicitly reported in their work. For L1455, we adopted the distance measured for L1451 instead, given the two clumps' proximity to each other on the sky. To calculate the mass per unit length (i.e., line mass; $M_{\mathrm{lin}}$) of each filament, we further divide each filament's mass by its spine length.

We note that, on the one hand, background emission and occasional filament overlap on the plane of the sky can lead to an overestimation of filament masses. On the other hand, \textit{Herschel}'s overestimation of dust temperatures for denser structures can lead to an underestimation of masses. \textit{Herschel}-derived dust temperature for the denser parts of B5, for example, is $\sim 3-4$ K higher on average than the NH$_3$ kinematic temperatures measured by \cite{Schmiedeke2021} at higher resolutions. Depending on the aggressiveness of the background removal, \citeauthor{Schmiedeke2021} estimated $M_{\mathrm{lin}}$ values for the B5 sub-filaments to be in the range of $40 - 80$, M$_\odot$ pc$^{-1}$, compared to 53 M$_\odot$ pc$^{-1}$ for our B5 filament. Considering that the B5 filament is composed of the two sub-filaments joined somewhat end-to-end, with a partial overlap in the longitudinal direction, the $M_{\mathrm{lin}}$ measurements in B5 should be relatively insensitive to spatial resolution. The consistency between the B5 $M_{\mathrm{lin}}$ estimates thus indicates that the systematic biases of our mass measurements tend to compensate for each other overall.

%============================================================
\section{Analyses \& Discussion} \label{sec:discussion}

%=============================
\subsection{The role of clump environment} \label{subsec:clump_env}

\begin{figure}[t]
\centering
\includegraphics[width=1\columnwidth]{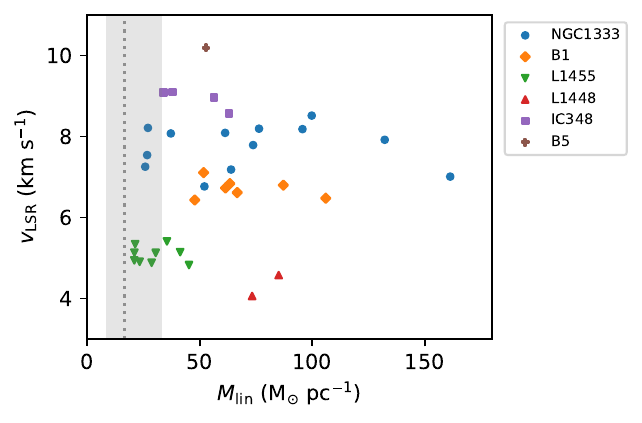}
\caption{The median $v_{\mathrm{LSR}}$ of each filament plotted against the filament's $M_{\mathrm{lin}}$. Filaments belonging to each star-forming clump are marked with symbols labeled in the legend. The value of $M_{\mathrm{lin, crit}}$ at 10 K is marked by the vertical dotted line, with the shaded region representing the transcritical values of $0.5 M_{\mathrm{lin, crit}} \leq M_{\mathrm{lin}} \leq 2 M_{\mathrm{lin, crit}}$. \label{fig:vlsr_v_lineMass}}
\end{figure}

As with molecular clouds in general, star formation in Perseus mostly occurs in higher-density clumps. In these locations, the probability density distribution of their column densities behaves like a power-law at N(H$_2$) $> 1 \times 10^{22}$ cm$^{-2}$, which breaks away from the log-normal-like behavior expected from their parental cloud \citep{Sadavoy2014}. For Perseus, we have NH$_3$ observations for the B5, IC 348, B1, NGC 1333, L1448, and L1455 star-forming clumps. To showcase clearly the $M_{\mathrm{lin}}$ and $v_{\mathrm{LSR}}$ ranges of the filaments within these clumps, Figure \ref{fig:vlsr_v_lineMass} plots these filaments' median $v_{\mathrm{LSR}}$ values against their respective $M_{\mathrm{lin}}$. Overall, the filaments within each Perseus clump are well separated from one another in the $v_{\mathrm{LSR}} - M_{\mathrm{lin}}$ space, mostly by $v_{\mathrm{LSR}}$. Interestingly, the spread of $M_{\mathrm{lin}}$ seems to vary from clump to clump, with NGC 1333 having a significantly larger spread than any other clump. Such a large spread suggests that filaments within a clump may not necessarily form at the same time, form the same way, or co-evolve in $M_{\mathrm{lin}}$ with each other.

Filaments with $M_{\mathrm{lin}}$ exceeding a critical value of $M_{\mathrm{lin, crit}} = 2c_s^2/G$ in an isothermal cylinder model in near hydrostatic equilibrium \citep{Ostriker1964} are expected to be unstable against radial contraction and longitudinal fragmentation \citep{Inutsuka1992}. In particular, radial collapse will dominate when filaments have $M_{\mathrm{lin}}$ significantly larger than $M_{\mathrm{lin, crit}}$. Here, the isothermal sound speed is calculated as $c_s = (k_{\mathrm{b}} T / \mu m_{\mathrm{H}})^{1/2}$, where $T$ is the isothermal gas temperature, and $G$, $k_{\mathrm{b}}$, $\mu$, and $m_{\mathrm{H}}$ are the gravitational constant, Boltzmann constant, mean interstellar molecular weight, and the atomic hydrogen mass, respectively. Assuming a mean molecular weight of $\mu = 2.33$, a 10 K filament would have $M_{\mathrm{lin, crit}} = 16.6$ M$_\odot$ pc$^{-1}$. We note that the 10 K temperature assumed here, commonly adopted in the literature, is consistent with those typically found in Perseus with NH$_3$ GBT observations ($\sim 11$ K; e.g., \citealt{Rosolowsky2008}; \citealt{Friesen2017}). 

As shown in Figure \ref{fig:vlsr_v_lineMass}, the filaments we identified in Perseus are all thermally supercritical, suggesting that they should either be contracting or fragmenting in the absence of additional support against gravity. While models show that radial collapse may quickly overwhelm fragmentation in filaments with $M_{\mathrm{lin}} \gtrsim 1.2 M_{\mathrm{lin, crit}}$ \citep{Inutsuka1997}, observations have shown that core formation likely occurs over a broader range of $M_{\mathrm{lin}}$ values (e.g., \citealt{Arzoumanian2013}, \citeyear{Arzoumanian2019}; \citealt{Konyves2015}, \citeyear{Konyves2020}). For example, the empirical core formation efficiency increases dramatically within a factor of 2 of the $M_{\mathrm{lin, crit}}$ value (i.e., $8-33$ M$_{\odot}$ pc$^{-1}$) before it plateaus at around $\sim 2M_{\mathrm{lin, crit}}$  (e.g., \citealt{Konyves2020}). Furthermore, cores in models can form in subcritical filaments with $M_{\mathrm{lin}} > 0.5M_{\mathrm{lin, crit}}$ due to compressive instabilities \citep{Fischera2012}. Therefore, following the convention of \cite{Arzoumanian2019}, we define a thermally transcritical regime as $0.5 M_{\mathrm{lin, crit}} \leq M_{\mathrm{lin}} \leq 2 M_{\mathrm{lin, crit}} $. By this definition, 9 (25\%) and 25 (69\%) of the 36 Perseus filaments in our sample are trans- and supercritical, respectively, assuming a gas temperature of 10 K. No filament is subcritical by the above definition and none of the transcritical filaments has $M_{\mathrm{lin}} < M_{\mathrm{lin, crit}}$, the classical critical value \citep{Ostriker1964}.

Most filaments identified by \cite{Pezzuto2021} in Perseus using the {\em Herschel} dust emission observations are thermally subcritical. This result, however, is not inconsistent with ours, given that NH$_3$ preferentially traces the densest filaments in Perseus. Indeed, most filaments identified by \cite{Pezzuto2021} are located well outside where N(H$_2$) $> 1 \times 10^{22}$ cm$^{-2}$ and where NH$_3$ emission are detected in our data. While \citeauthor{Pezzuto2021} have also identified filaments in regions that overlap with our data, many of the shorter filaments identified here are missing from their sample, particularly in crowded environments like NGC 1333. Moreover, filaments from \citeauthor{Pezzuto2021} that overlap with our sample tend to extend further to the lower column density regions where NH$_3$ are not detected. This bias would contribute additionally to why the $M_{\mathrm{lin}}$ calculated by \citeauthor{Pezzuto2021} tends to be lower even for ``denser'' filament samples.

\begin{figure}[t]
\centering
\includegraphics[width=0.955\columnwidth]{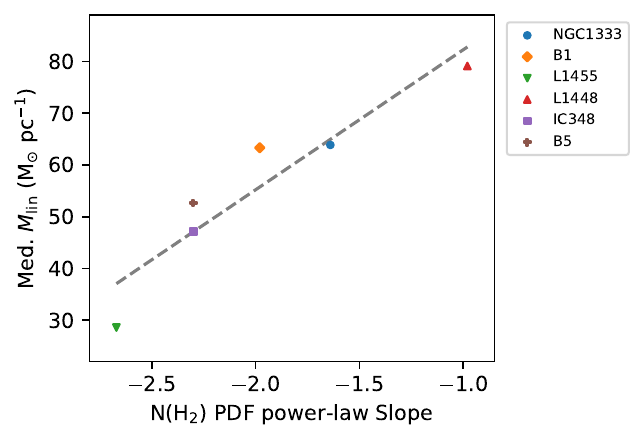}
\caption{The median $M_{\mathrm{lin}}$ of filaments within each star-forming clump plotted against the N(H$_2$) PDF power-law slope of the clump measured by \cite{Pezzuto2021} using the same method as \cite{Sadavoy2014}. The symbols are the same as those in Figure \ref{fig:vlsr_v_lineMass}. The grey dashed line marks the best-fit linear regression model. \label{fig:pdfSlope_v_mline}}
\end{figure}

The mass of each Perseus clump summed over areas with N(H$_2$) $ \geq 1 \times 10^{22}$ cm$^{-2}$ are 365 M$_{\odot}$, 342 M$_{\odot}$, 101 M$_{\odot}$, 118 M$_{\odot}$, 156 M$_{\odot}$, and 28 M$_{\odot}$ for NGC 1333, B1, L1455, L1448, IC 348, and B5, respectively \citep{Sadavoy2014}. We find no correlation between the Perseus filaments' median $M_{\mathrm{lin}}$ within a clump and their respective clump mass. In particular, L1448 and L1455 have very comparable clump masses (within  20\%) but L1448's median $M_{\mathrm{lin}}$ is about three times that of L1455's. The clumps' power-law index of N(H$_2$) probability density function (PDF) above N(H$_2$) $\sim 1 \times 10^{22}$ cm $^{-2}$ (\citealt{Sadavoy2014}; \citealt{Pezzuto2021}), on the other hand, correlates well with the median $M_{\mathrm{lin}}$ of a clump. Figure \ref{fig:pdfSlope_v_mline} shows the tight relations between these two clump properties, with a Pearson's $r$ value of 0.94 and a best-fit linear regression model slope of $27 \pm 5$ M$_{\odot}$ pc$^{-1}$. Evidently, the increased concentration of mass within a clump, as indicated by the flatter PDF slope, directly enhances the $M_{\mathrm{lin}}$ of the filaments within. This result has an important implication given that PDF slopes correlate with SFE both in simulations \citep{Federrath2013} and in observations (e.g., Perseus; \citealt{Sadavoy2014}).

\begin{figure}[t]
\centering
\includegraphics[width=0.955\columnwidth]{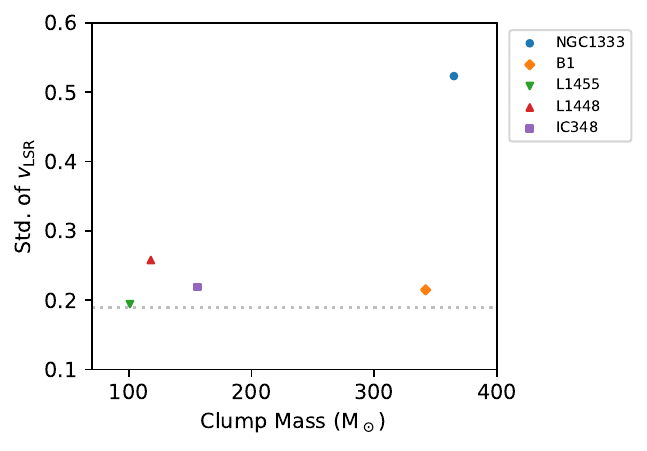}
\caption{The inter-filament standard deviation of each filament's median $v_{\mathrm{LSR}}$ within a star-forming clump plotted against the mass of the clump above N(H$_2$) $= 1\times 10^{22}$ cm$^{-1}$, as measured by \cite{Sadavoy2014}. The symbols are the same as those in Figure \ref{fig:vlsr_v_lineMass}. The horizontal dotted line marks the isothermal sound speed of a 10 K gas. \label{fig:vlsrSpread_v_clumpMass}}
\end{figure}

To understand better the relative velocities of filaments within a clump, Figure \ref{fig:vlsrSpread_v_clumpMass} shows the standard deviation of filaments' median $v_{\mathrm{LSR}}$ in each clump plotted against the clump's total mass, with B5 excluded due to its limited sample size of one. As demonstrated, the two properties do not correlate and, except for NGC 1333, all seem to have inter-filament $v_{\mathrm{LSR}}$ dispersion of $\sim 0.22$ km s$^{-1}$ regardless of the clump mass. Moreover, given this near-constant dispersion, inter-filament $v_{\mathrm{LSR}}$ dispersion correlates with neither the average clump densities nor the PDF power-law slopes. Interestingly, this near-constant dispersion value is just above the sound speed of a 10 K gas. Filaments in these clumps may thus originate from a shared sonic process, such as a larger-scale convergence flow that becomes impressible and consequently stagnant at the sonic scale.

While NGC 1333 has roughly the same clump mass as B1, the former's inter-filament $v_{\mathrm{LSR}}$ dispersion is $\sim 2.5$ times that of the latter and other Perseus clumps. Such a difference indicates that clumps above a certain mass threshold may start to evolve dynamically due to their larger gravitational potential. NGC 1333 and B1, for example, both have clump masses more than twice that of the other Perseus clumps, but NGC 1333 has $\sim 23\%$ higher average column density and a flatter PDF power-law slope, indicating that NGC 1333 is more evolved than B1. Indeed, NGC 1333 is the only clump in Perseus with a significant NH$_3$ velocity gradient on the clump scale, which suggests that NGC 1333 is actively accreting from its surroundings \citep{ChenM2020}. In this case, already-formed filaments may fall into this larger-scale potential to produce the higher inter-filament velocity dispersion observed, similar to how gravitational contraction gathers filaments in simulations by \cite{Smith2016}.

Alternatively, the filaments observed in clumps like NGC 1333 may be produced in situ from accretion-driven turbulence on the clump scale similar to that proposed by \cite{Matzner2015}. Since NGC 1333 contains significantly more filaments than B1 despite their comparable clump masses, this latter scenario where filaments are produced in situ is thus more likely. NGC 1333's higher SFE and flatter PDF power-law slope \citep{Sadavoy2014}, as well as larger dust grain population \citep{ChenMCY2016}, further support that NGC 1333 is more evolved than B1.

Even though stellar feedback in NGC 1333 is more significant than the rest of Perseus, they are unlikely a primary driver behind NGC 1333's higher inter-filament $v_\mathrm{LSR}$ dispersion. Protostellar outflows, for example, tend to inject kinetic energy on the smaller scales and cannot drive bulk filament motions that result in the observed inter-filament dispersion. While a bubble appears to be expanding into NGC 1333 (e.g., \citealt{Dhabal2019}, \citealt{ChenMike2024}), its contact appears fairly limited and localized, which cannot solely explain the large $v_\mathrm{LSR}$ dispersion across NGC 1333 either. The elevated inter-filament $v_\mathrm{LSR}$ dispersion in NGC 1333 is thus unlikely driven by local stellar feedback. We briefly discuss the role of such feedback on the smaller scales in Section \ref{subsec:cores}.

%======================================================
\subsection{Internal velocity dispersion of filaments} \label{subsec:sigv_n_vgrad}

\begin{figure}[t]
\centering
\includegraphics[width=1.0\columnwidth]{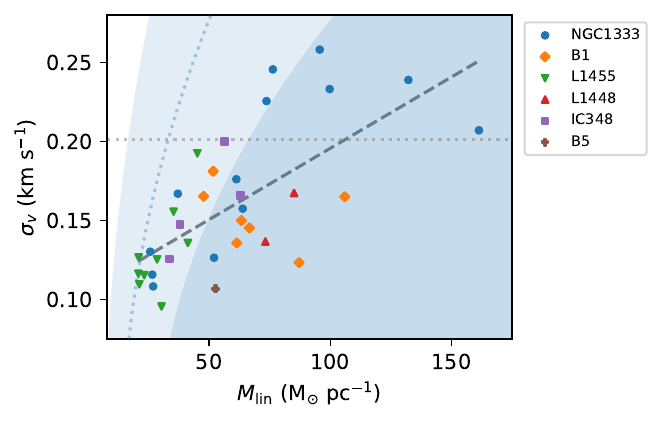}
\caption{The median $\sigma_v$ of each filament plotted against the filament's $M_{\mathrm{lin}}$, with the same markers as those in Figure \ref{fig:vlsr_v_lineMass}. The grey dashed line marks the best-fit linear regression model. The light and dark blue shaded regions show where $0.5 M_{\mathrm{line, vir}} \leq M_{\mathrm{line}} \leq 2 M_{\mathrm{line, vir}}$ and $M_{\mathrm{line}} \geq 2 M_{\mathrm{line, vir}}$, respectively. The dotted line marks where $M_{\mathrm{line, vir}} = M_{\mathrm{line}}$. The observed $\sigma_{\mathrm{v}}$ contains both a thermal and a non-thermal component, and the horizontal dotted line shows the $\sigma_{\mathrm{v}}$ value of a 10 K gas with a non-thermal component that is sonic.
\label{fig:sigv_v_lineMass}}
\end{figure}

Accreting filaments can have higher $\sigma_v$ than their non-accreting counterparts due to accretion-driven turbulence (e.g., \citealt{Klessen2010}). Figure \ref{fig:sigv_v_lineMass} shows the median $\sigma_v$ of each filament plotted against their respective $M_{\mathrm{lin}}$. Similar to the results found by \cite{Arzoumanian2013} for virially-bound filaments, the $\sigma_v$ and $M_{\mathrm{lin}}$ of the Perseus filaments are correlated. Specifically, these data have a Pearson's correlation coefficient (Pearson's $r$) of 0.68 and a best-fit linear regression model slope of $0.9 \pm 0.2$ m s$^{-1}$ M${_\odot}^{-1}$ pc. For comparison, \cite{Arzoumanian2013} fitted their data using a power-law model and obtained a best-fit power-law slope index of $0.36 \pm 0.07$. The linear slope approximation of their power-law over our $M_{\mathrm{lin}}$ range is consistent with our best-fit slope, but their model's $\sigma_v$ appears systematically higher than ours by $\sim 1.5$ km s$^{-1}$.

Given nearly all the bound filaments measured by \citeauthor{Arzoumanian2013} are drawn from N$_2$H$^+$ observations, which traces denser gas than NH$_3$ and is expected to have lower $\sigma_v$ due to lower volume-filling fraction, tracer difference alone cannot account for such a difference. This $\sigma_v$ difference also seems too large to be accounted for by spatial resolution discrepancy either, which suggests environmental differences are at play. For example, a lower $\sigma_v$ may result from lower star formation activities that inject turbulence into the cloud via feedback. Similarly, a lower $\sigma_v$ can also result from a stronger local magnetic field that better supports filaments against self-gravity while damping the turbulence within. Such magnetic support can also suppress star formation within a cloud and further reduce the turbulence from stellar feedback. While increases in gas temperatures from stellar feedback can also increase $\sigma_v$, the filaments sampled by \citeauthor{Arzoumanian2013} and in this study all seem to be reasonably isothermal, with temperatures around 10 K.

\cite{Hacar2023} also found a correlation between $\sigma_v$ and $M_{\mathrm{lin}}$ using a large conglomerate of literature data that spans four orders of magnitude in $M_{\mathrm{lin}}$. Many of their smaller filament samples are hierarchically embedded in larger filaments, and many of their larger filament samples are not sufficiently resolved to detect substructures. \citeauthor{Hacar2023} reported a best-fit power-law index of 0.5. Their model, however, is only derived from filaments with $M_{\mathrm{lin}} > 100$ M${_\odot}^{-1}$ pc. The scatter in their data is also quite large, likely inherited from the different systematic biases of such a large and diverse set of observations.

The observed correlation between $\sigma_v$ and $M_{\mathrm{lin}}$ has been interpreted by \cite{Arzoumanian2013} as the consequences of accretion/contraction driven turbulence (e.g., \citealt{Klessen2010}) where turbulence increases with filament growth via accretion. Indeed, the filament accretion rate does appear to depend on $M_{\mathrm{lin}}$ in simple analytical models, where $M_{\mathrm{lin}}$ grows over time and subsequently increases accretion-driven turbulence \citep{Heitsch2013}. This interpretation also explains why \citeauthor{Arzoumanian2013} did not find a similar correlation for unbound and thermally subcritical filaments, given that such filaments are not expected to be contracting. While spatially unresolved velocity gradients (e.g., from bulk motions) can contribute to $\sigma_v$ in addition to turbulence, we estimated the median contribution of from such a source to be $\sim 0.05$ km s$^{-1}$ using a $6''$-resolution NH$_3$ data \citep[see][]{ChenMike2024}, which is not particularly significant.

Since non-thermal gas motions could, in principle, provide additional pressure support to filaments, a virial mass per unit length, i.e., $M_{\mathrm{line, vir}} = 2 \sigma_{\mathrm{v}}^2/G$, can be calculated to account for such additional support \citep{Fiege2000}. We note that $\sigma_{\mathrm{v}}$ is directly measured from our line fits and encompasses both the thermal and non-thermal motions, i.e., $\sigma_{\mathrm{v}}^2 = \sigma_{\mathrm{v,T}}^2 + \sigma_{\mathrm{v,NT}}^2$, and that such support is an upper limit estimate. Depending on the model, not even turbulence can necessarily provide filaments with pressure support \citep{Heigl2020}. Figure \ref{fig:sigv_v_lineMass} shows where $0.5 M_{\mathrm{line, vir}} \leq M_{\mathrm{line}} \leq 2 M_{\mathrm{line, vir}}$ and $M_{\mathrm{line}} \geq 2 M_{\mathrm{line, vir}}$ in the light and dark blue-shaded regions, respectively. The former and the latter regimes can be considered the virial trans- and supercritical counterparts of the thermal masses per unit length. For $M_{\mathrm{line}} \leq 0.5 M_{\mathrm{line, vir}}$, the filament is considered gravitationally unbound.

\begin{figure}[t]
\centering
\includegraphics[width=1.0\columnwidth]{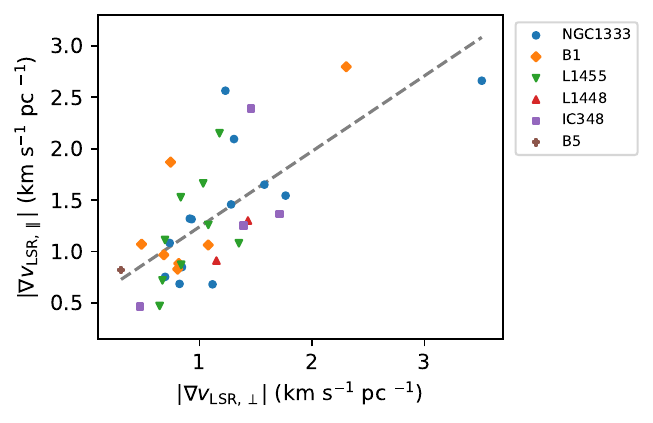}
\caption{The median $|\nabla v_{\mathrm{LSR}, \perp}|$ of the Perseus filaments plotted against their $|\nabla v_{\mathrm{LSR}, \parallel}|$ counterparts. The markers are the same as those found in Figure \ref{fig:vlsr_v_lineMass}. The best-fitted linear regression models are represented by the grey dashed line. \label{fig:vgrad_perp_vs_para}} %The solid-shaded region shows the area between the 2:1 and 1:2 lines
\end{figure}

Consistent with findings of \cite{Arzoumanian2013}, all the Perseus filaments are gravitationally bound by the virial criterion (i.e., $M_{\mathrm{line}} > 0.5 M_{\mathrm{line, vir}}$). Out of the 36 Perseus filaments, 20 (56\%) of them are transcritical, and 16 (44\%) of them are supercritical. Moreover, 35 (97\%) of all filaments have $M_{\mathrm{line}} > M_{\mathrm{line, vir}}$, the classical virially critical value against radial collapse \citep{Fiege2000}. These results suggest that most filaments in our sample are unlikely to be in equilibrium even if non-thermal motions can provide significant additional support against self-gravity. Unless there is a substantial amount of magnetic support as well, most of these filaments are likely accreting actively from their surroundings to sustain themselves from a total radial collapse as proposed by \cite{Arzoumanian2013}. 

If thermally trans- and super-critical filaments do rely on continuous accretion to survive and form cores, then a filament's $M_{\mathrm{line}}$ can serve as an evolution indicator of its mass growth. Considering that the clump with the lowest $M_{\mathrm{lin}}$ values in Perseus (L1455) has the lowest fraction of thermally supercritical filaments and the lowest star formation efficiency (SFE) \citep{Sadavoy2014}, filaments may indeed be formed as thermally sub- or transcritical entities and subsequently grow through mass accretion and becoming supercritical.

%%%%%%%%%%%%%%%%%%%%%%%%%%%%%%%%%%%%%%%%%%%%%%%%%%%%%%%%%%%%%
\subsection{Internal velocity gradients of filaments} \label{subsec:vgrad}

\begin{figure*}[t!]
\centering
\includegraphics[width=1.0\textwidth, trim={-11mm 0 0 0}]{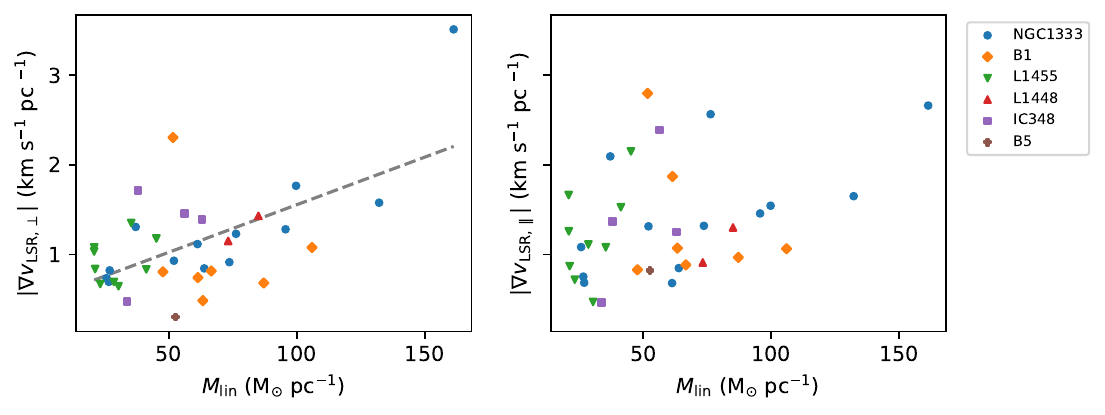}
\includegraphics[width=1.0\textwidth, trim={-11mm 0 0 0}]{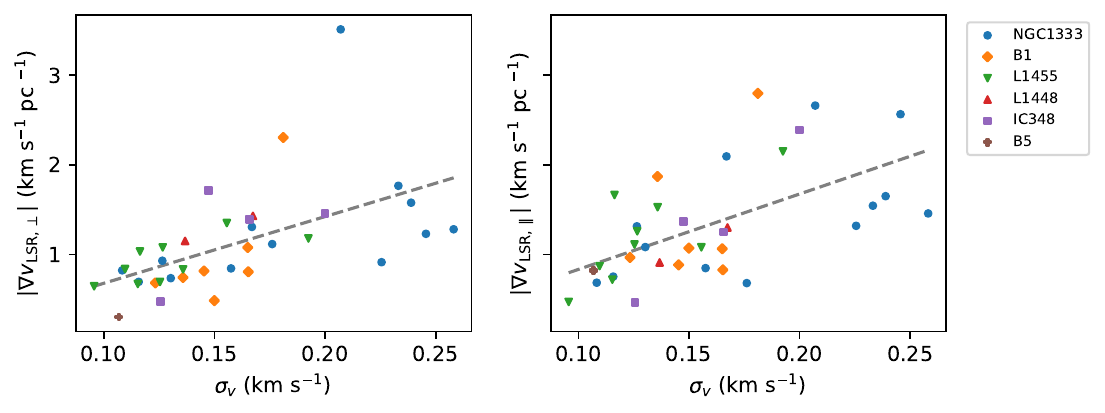}
\caption{The the median $|\nabla v_{\mathrm{LSR}, \perp}|$ (left column) and the $|\nabla v_{\mathrm{LSR}, \parallel}|$ (right column) of the Perseus filament plotted against their $M_{\mathrm{lin}}$ and $\sigma_v$ counterparts in the top and bottom rows, respectively. We note that $\sigma_v$ is the line-of-sight velocity dispersion derived directly from the spectral fits. The markers are the same as those found in Figure \ref{fig:vlsr_v_lineMass}. The grey dashed line represents the best-fitted linear regression models. \label{fig:vgrad_vs_lineMass_n_sig}}
\end{figure*}

Observed velocity gradients that are perpendicular (e.g., \citealt{Fernandez-Lopez2014}) and parallel (e.g., \citealt{KirkHelen2013}) relative to the filament spine have often been interpreted as mass flows onto and along filaments, respectively. Figure \ref{fig:vgrad_perp_vs_para} shows the median magnitudes of the perpendicular component of $\nabla v_{\mathrm{LSR}}$ (i.e., $|\nabla v_{\mathrm{LSR}, \perp}|$), plotted against their parallel counterparts (i.e., $|\nabla v_{\mathrm{LSR}, \parallel}|$) for individual Perseus filaments. The data have a Pearson's $r$ value of 0.70 and best-fit linear-regression slope of $0.73 \pm 0.13$. 

The top row of Figure \ref{fig:vgrad_vs_lineMass_n_sig} shows the median $|\nabla v_{\mathrm{LSR}, \perp}|$ (left) and the $|\nabla v_{\mathrm{LSR}, \parallel}|$ (right) of the Perseus filament plotted against their respective $M_{\mathrm{lin}}$. Interestingly, $M_{\mathrm{lin}}$ correlates with $|\nabla v_{\mathrm{LSR}, \perp}|$ but not necessarily $|\nabla v_{\mathrm{LSR}, \parallel}|$. For example, running the Wald Test with a t-distribution shows that the former and latter have $p$-values of $2 \times 10^{-4}$ and 0.03, respectively, where the $p$-value corresponds to the null-hypothesis probability that the observed data can be drawn from a zero-slope distribution. Moreover, Pearson's $r$ values for the former and the latter are 0.59 and 0.37, respectively. For this reason, we only fit a linear regression model to the former and not the latter. The model that best fit the $|\nabla v_{\mathrm{LSR}, \perp}| - M_{\mathrm{lin}}$ relation has a slope of  $11 \pm 3$ m s$^{-1}$ M$_\odot$.

Since $\nabla v_{\mathrm{LSR}}$ can be a sign of acceleration/deceleration associated with gas flow, the correlation between $|\nabla v_{\mathrm{LSR}, \perp}|$ and $M_{\mathrm{lin}}$ likely indicates that the radial (i.e., perpendicular) accretion rate of filaments does indeed increase with $M_{\mathrm{lin}}$, much in the manner predicted by analytical models (e.g., \citealt{Heitsch2013}). Even if the observed $\nabla v_{\mathrm{LSR}}$ does not directly trace accretion flow, inhomogeneous radial accretion can still produce prominent and localized $|\nabla v_{\mathrm{LSR}, \perp}|$ by generating vortices (macroscopic turbulence) with axes running parallel to filament spines (e.g., \citealt{Clarke2017}). Indeed, the morphology of clumpy, alternating $\nabla v_{\mathrm{LSR}, \perp}$ structures found by \cite{ChenM2020} in NGC 1333 using the same NH$_3$ data here and the elongated $\nabla v_{\mathrm{LSR}, \perp}$ structures found by \cite{ChenMCY2022} in B5 using higher resolution ($\sim 0.01$ pc) NH$_3$ data are qualitatively consistent with the line-of-sight velocity structures seen in simulations.

The $\nabla v_{\mathrm{LSR}, \parallel}$ values seen on the filament-length scales, on the other hand, have often been attributed to longitudinal mass flow accreting towards a star-forming hub (e.g., \citealt{KirkHelen2013}). Under such a scenario, $|\nabla v_{\mathrm{LSR}, \parallel}|$ is expected to be driven by the gravitational potential of the hub rather than $M_{\mathrm{lin}}$. Such a scenario may explain why $|\nabla v_{\mathrm{LSR}, \parallel}|$ and $M_{\mathrm{lin}}$ are not well correlated in Perseus. Since the Perseus clumps do not feature prominent star-forming hubs (with the notable exception of NGC 1333), we do not necessarily expect $|\nabla v_{\mathrm{LSR}, \parallel}|$ to depend on $M_{\mathrm{lin}}$ indirectly via co-evolution of hubs and filaments either.

\begin{figure*}[t!]
\centering
\includegraphics[width=1.0\textwidth, trim={-11mm 0 0 0}]{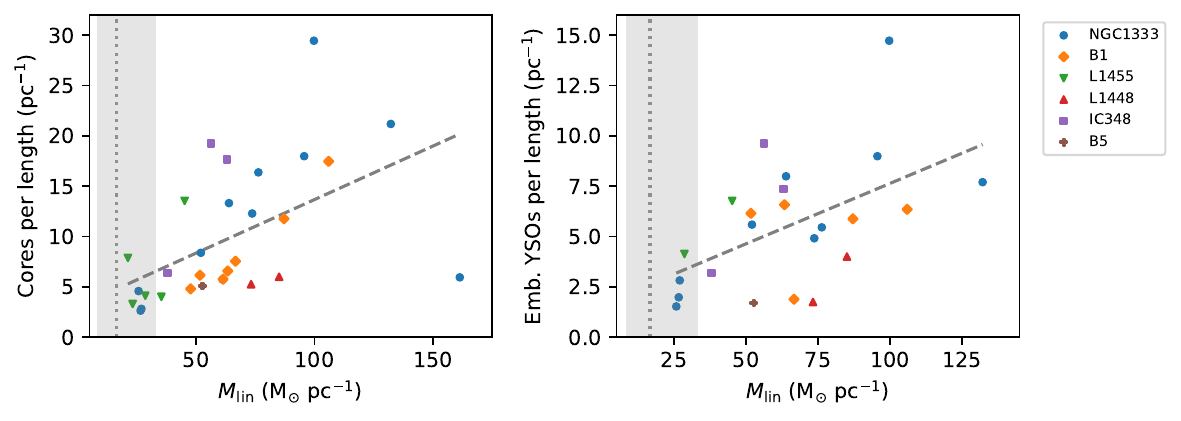}
\includegraphics[width=1.0\textwidth, trim={-11mm 0 0 0}]{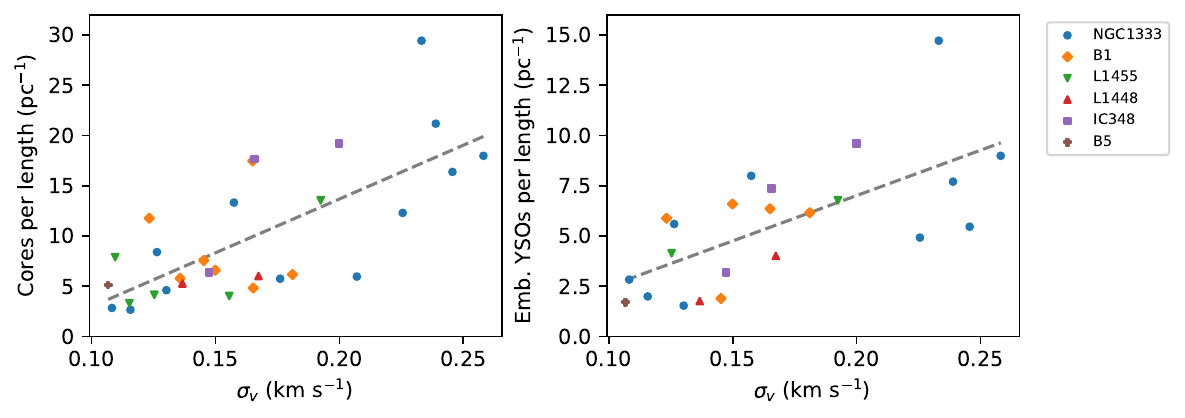}
\caption{The number of cores per unit length (left column) and the number of class 0/1 YSOs per unit length (right column) of Perseus filaments plotted against their $M_{\mathrm{lin}}$ and median $\sigma_v$ counterparts in the top and bottom rows, respectively. The symbols are the same as those found in Figure \ref{fig:vlsr_v_lineMass}. The grey dashed line represents the best-fitted linear regression model. The value of $M_{\mathrm{lin, crit}}$ at 10 K is marked by the vertical dotted line in the top panels, with shaded region representing the transcritical values of $0.5 M_{\mathrm{lin, crit}} \leq M_{\mathrm{lin}} \leq 2 M_{\mathrm{lin, crit}}$. \label{fig:cores_n_YSOs_PerL_vs_lineMass_n_sig}}
\end{figure*}

Considering the $\nabla v_{\mathrm{LSR}}$ vectors we measured are on scales much smaller than the filament lengths (e.g., $\sim 0.05$ pc), the $\nabla v_{\mathrm{LSR}, \parallel}$ values we find could also be probing other kinematic structures, such as longitudinal filament fragmentation (e.g., \citealt{Inutsuka1992}), transversal oscillation (e.g., \citealt{Gritschneder2017}), magnetosonic waves (e.g., \citealt{Heyer2016}), or shell instability \citep{Anathpindika2022}. If these processes are indeed responsible for the observed $\nabla v_{\mathrm{LSR}, \parallel}$ behavior found in Perseus, then they do not seem to be strongly coupled to either the $M_{\mathrm{lin}}$ of the filament or other filament properties that co-evolve with $M_{\mathrm{lin}}$.

To examine the role of accretion-driven turbulence further, Figure \ref{fig:vgrad_vs_lineMass_n_sig}'s bottom row plots the median $|\nabla v_{\mathrm{LSR}, \perp}|$ (left) and the $|\nabla v_{\mathrm{LSR}, \parallel}|$ (right) of Perseus filaments against their respective median $\sigma_v$ values. The results show each of the perpendicular and parallel $|\nabla v_{\mathrm{LSR}}|$ components correlate with $\sigma_v$ and have Pearson's $r$ values of 0.55 and 0.59, respectively. The best-fit linear regression slopes of these trends are $7.4 \pm 2$ pc$^{-1}$ and $8.4 \pm 2$ pc$^{-1}$, respectively. The correlation between $|\nabla v_{\mathrm{LSR}}|$ and $\sigma_v$ does not discriminate strongly between their perpendicular and parallel components. Given that $|\nabla v_{\mathrm{LSR}, \parallel}|$ correlate with $\sigma_v$ but not with $M_{\mathrm{lin}}$, the increase in $\sigma_v$ with $M_{\mathrm{lin}}$ (Figure \ref{fig:sigv_v_lineMass}) is unlikely caused solely by an increase in radial accretion driven by higher $M_{\mathrm{lin}}$. Instead, other drivers that increase $|\nabla v_{\mathrm{LSR}, \parallel}|$, such as fragmentation, are likely more responsible for increasing $\sigma_v$. We will revisit this speculation with respect to core formation in Section \ref{subsec:cores}.

%%%%%%%%%%%%%%%%%%%%%%%%%%%%%%%%%%%%%%%%%%%%%%%%%%%%%%%%%%%%%%%%
\subsection{Filament fragmentation, cores, and YSOs} \label{subsec:cores}

Cores have been proposed to form in thermally supercritical filaments via fragmentation \citep{Andre2014}. In quasi-equilibrium (e.g., \citealt{Inutsuka1992}) and non-equilibrium (e.g., \citealt{Clarke2016}) models of isolated filaments, such fragmentation is seeded during the subcritical phase by the fastest growing perturbation. In these models, the number of cores within each filament is also not expected to increase once supercriticality is reached because further perturbation growth would be outpaced by existing filament fragmentation or radial collapse. As discussed earlier in Section \ref{subsec:sigv_n_vgrad}, fragmentation that takes place beyond the subcritical phase may occur over a smoother, transcritical regime of $0.5 M_{\mathrm{lin, crit}} \leq M_{\mathrm{lin}} \leq 2 M_{\mathrm{lin, crit}}$ (e.g., \citealt{Fischera2012}; \citealt{Konyves2015}, \citeyear{Konyves2020}) rather than a sharp boundary (e.g., \citealt{Ostriker1964}).

Figure \ref{fig:cores_n_YSOs_PerL_vs_lineMass_n_sig}'s top left panel shows the number of cores per unit length in each Perseus filament plotted against their $M_{\mathrm{lin}}$ counterparts. We use the cores from the JCMT GBS core catalog by Pattle et al. (submitted; see also Figure \ref{fig:Spine_Maps} for positions) and define the filament lengths by the extent of their spines in the sky. We note that this approach ignores the foreshortening of the filaments by inclination but is systematically consistent with how we calculated $M_{\mathrm{lin}}$. The data have a Pearson's $r$ value of 0.52 and a best-fit linear regression slope in the number of cores per M${_\odot}$ of $0.11 \pm 0.01$. This correlation suggests that $M_{\mathrm{lin}}$ and the average core spacing in a filament are physically related. 

The aforementioned near-equilibrium models, however, do not predict the number of cores in a given filament to co-evolve with $M_{\mathrm{lin}}$ beyond the subcritical phase. The filament fragmentation seeded during the sub- and transcritical phases will outpace any new perturbations beyond these phases and prevent new cores from forming. Since none of our filaments are thermally subcritical, the filament growth rate (and hence $M_{\mathrm{lin}}$) during the supercritical phase would have to be pre-determined by the initial conditions that also set the core fragmentation spacing. Inhomogeneous accretion from a highly turbulent environment, for example, could simultaneously produce more filament substructures (e.g., \citealt{Clarke2017}) and drive $M_{\mathrm{lin}}$ towards higher values during their subsequent evolution. 

Alternatively, the correlation between the number of cores per unit length and $M_{\mathrm{lin}}$ may suggest the number of cores can continue to increase as filaments evolve well into the supercritical stage. Indeed, simulated filaments embedded within a more realistic cloud environment, rather than existing in isolation, have shown number of fragments (i.e., cores) and mean $M_{\mathrm{lin}}$ that increase consistently over time, even well after their mean $M_{\mathrm{lin}}$ is significantly above the supercritical value \citep{Chira2018}. Moreover, the typical separation between these fragments and their closest neighbors within the same filament decreases over time, likely driven predominately by the emergence of new fragments. These results imply that the cores-per-unit-length values of these simulated filaments correlate with their respective $M_{\mathrm{lin}}$. Such a co-evolution can well explain the observed correlation between cores per unit length and $M_{\mathrm{lin}}$ in the Perseus filaments.

While length contraction is commonly found in filament models, their corresponding collapse time scales are substantially longer than the filament perturbation (i.e., fragmentation) time scale, particularly for higher line masses \citep[e.g.,][]{Seifried2015, Hoemann2023}. In fact, higher line mass filaments do not typically contract significantly by the time a majority of sink particles have formed \citep{Seifried2015}. Moreover, filament accretion can slow down such a process \citep{Heigl2022}, even maintaining or increasing filament lengths \citep{Feng2024}. We, therefore, do not expect length contraction to play a significant role in increasing cores per unit length in our study.

To see if the co-evolution between cores and $M_{\mathrm{lin}}$ also maps onto newly-formed, deeply embedded protostars, we plotted the number of Class 0/I YSOs per unit length in each Perseus filament plotted against their $M_{\mathrm{lin}}$ counterparts in Figure \ref{fig:cores_n_YSOs_PerL_vs_lineMass_n_sig}'s top right panel. We take our Class 0/I YSO samples from the Spitzer c2d catalog by \cite{Dunham2015} and select only those found within the on-sky footprints of each filament. We further impose that only one YSO is counted per beam to decontaminate intra-core stellar multiplicities and enable our YSO counts as a proxy for the protostellar core counts. The positions of these YSOs are shown in Figure \ref{fig:Spine_Maps}. The best-fit linear regression to this data has a slope of $0.06 \pm 0.02$ YSOs per M${_\odot}$ and the data has a Pearson's $r$ value and a corresponding p-value for a zero-slope null hypothesis of 0.53 and 0.01, respectively. This p-value is less robust than the other correlations we found due to smaller number statistics. Given that cores are direct progenitors of YSOs, and the Class 0/I YSOs in our samples are expected to be individually embedded in protostellar core, the correlation we find here is likely indeed driven by a simple mapping from cores per unit length to YSOs per unit length.

Figure \ref{fig:cores_n_YSOs_PerL_vs_lineMass_n_sig}'s bottom row shows the number of cores per unit length (left) and their Class 0/I YSOs counterparts (right) for each Perseus filament plotted against their respective median $\sigma_v$ values. These two properties have Pearson's $r$ values of 0.71 and 0.67, and best-fit linear regression slopes in numbers per pc per km s$^{-1}$ of $107 \pm 20$ and $45 \pm 11$, respectively. As discussed in Section \ref{subsec:vgrad}, the correlation between $\sigma_v$ and $M_{\mathrm{lin}}$ in Perseus filaments (Figure \ref{fig:sigv_v_lineMass}) may not be driven purely by higher radial accretion rates of filaments with larger $M_{\mathrm{lin}}$ values. The correlation between $M_{\mathrm{lin}}$ and cores or YSOs per unit length may thus suggest infall along filaments towards cores or YSOs can also contribute significantly to the increase of $\sigma_v$ in filaments, in addition to radial accretion onto filaments. Indeed, N$_2$H$^+$ observations of Perseus cores have shown that protostellar cores, which are collapsing, have $\sigma_v$ values that are on average $\sim 0.5$ km s$^{-1}$ higher than their prestellar counterparts \citep{KirkHelen2007}. NH$_3$ observations of Perseus cores also found a similar but less pronounced result \citep{Foster2009}. The latter study attributes the differences between these two studies to differences in core identification rather than the tracer biases.

Even though protostellar outflows in Perseus can inject a significant amount of turbulence into the cloud as well (e.g., \citealt{Arce2010}; \citealt{XuDuo2020}), we do not see localized $\sigma_v$ enhancement that well correlates with outflows identified by \cite{Stephens2017}. This lack of correlation suggests that outflows do not impact dense gas kinematics significantly or directly. Indeed, \cite{Foster2009} did not find a difference between the isolated and clustered cores' non-thermal linewidths in NH$_3$. Moreover, given \citeauthor{Stephens2017} cannot rule out that outflows in Perseus are randomly oriented relative to their host filaments, these outflows are more likely to impact their ambient surroundings rather than their hosts. Thus, even if outflows do inject a significant amount of turbulent energy into the denser gas, they are more likely to increase the $\sigma_v$ in their surroundings than their parental filaments. The observed correlation between $\sigma_v$ and the number of cores or YSOs per unit length is thus unlikely driven by stellar feedback from within the hosting filaments. In fact, the injection of turbulent energy via outflows may introduce additional scatter to such a trend in highly clustered star-forming environments, such as NGC 1333, where filaments of any kind can be impacted by outflows launched from neighboring filaments.

\begin{figure}[t]
\centering
\includegraphics[width=1.0\columnwidth]{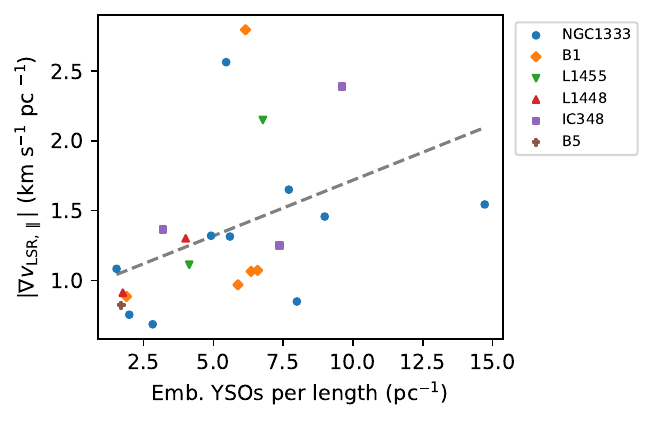}
\caption{The median $|\nabla v_{\mathrm{LSR}, \perp}|$ of each filament plotted against their respective number of Class 0/I YSOs. The markers are the same as those in Figure \ref{fig:vlsr_v_lineMass}. The grey dashed line represents the best-fitted linear regression model. \label{fig:vgradPara_vs_YSOpL}}
\end{figure}

While we do not find $|\nabla v_{\mathrm{LSR}, \parallel}|$ and the number of cores per unit length to correlate, Figure \ref{fig:vgradPara_vs_YSOpL} shows that such a correlation exists mildly between $|\nabla v_{\mathrm{LSR}, \parallel}|$ and the number of Class 0/I YSOs per unit length. These two latter properties have a Pearson's $r$ value of 0.43 and a best-fit linear regression slope of $0.08 \pm 0.04$ km s$^{-1}$. With a $p$-value of 0.04, however, we cannot confidently rule out a zero-slope null hypothesis, partially due to the smaller sample size. Nevertheless, if this correlation is indeed real, then the increased $|\nabla v_{\mathrm{LSR}, \parallel}|$ in filaments on the $0.05$ pc scale is likely driven by longitudinal infall towards the embedded YSOs or (magneto)hydrodynamic waves excited by such a process. The fact no such a correlation is found between the number of cores per unit length and $|\nabla v_{\mathrm{LSR}, \parallel}|$ further suggests the formation of prestellar cores via filament fragmentation does not produce a significant amount of $|\nabla v_{\mathrm{LSR}, \parallel}|$ until YSOs start to form later in the evolution. The lack of evidence of outflows directly impacting the host filaments further suggests the correlation between embedded YSOs per unit length and $|\nabla v_{\mathrm{LSR}, \parallel}|$ are not driven by protostellar feedback. 

%=============================
\subsection{Velocity gradient orientations} \label{subsec:vgrad_orientation}

A preferential, non-random $\nabla v_{\mathrm{LSR}}$ orientation on the plane-of-sky can indicate a well-ordered acceleration field within a filament, likely governed by gravity, magnetic fields, or large-scale convergence flow. To see whether or not the $\nabla v_{\mathrm{LSR}}$ vectors are randomly oriented relative to each other within each filament, we performed the Rayleigh test of uniformity \citep{Wilkie83}. The test adopts a null hypothesis that assumes the vector orientations to be uniformly distributed around a circle and an alternative hypothesis that assumes the converse. Of the 36 Perseus filaments, only 4 (11\%) have Rayleigh $p$-values less than 0.001. This result indicates that 89\% of the Perseus filaments have highly non-random $\nabla v_{\mathrm{LSR}}$ orientations, consistent with the findings of \cite{ChenM2020} in NGC 1333 using the same data here. 

Even though the Perseus filaments' $\nabla v_{\mathrm{LSR}}$ orientations are highly non-random, they are not typically distributed unimodally either. Many of these distributions appear bimodal or multimodal by eye, with no discernible pattern on how the individual modes are distributed relative to each other (see also \citealt{ChenM2020}). For these reasons, we do not include circular means or circular standard deviation in our analysis. Nevertheless, the Rayleigh test $p$-values alone can still indicate how non-random the $\nabla v_{\mathrm{LSR}}$ orientations are without these metrics.

\begin{figure}[t]
\centering
\includegraphics[width=1.0\columnwidth]{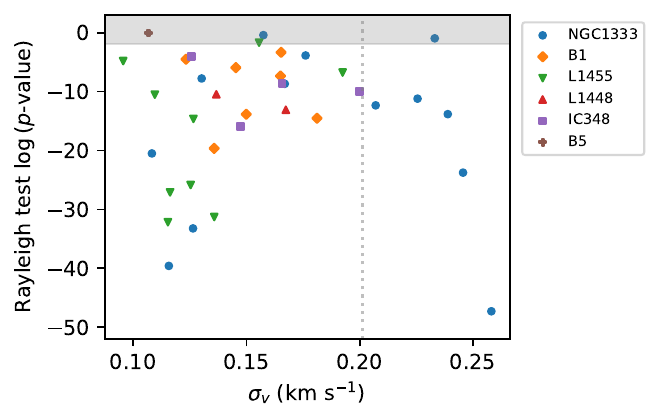}
\caption{The Rayleigh test $p$-values for each filament's $\nabla v_{\mathrm{LSR}}$ orientations plotted against their respective median $\sigma_v$, with the null hypothesis assuming the orientations are randomly distributed. The markers are the same as those in Figure \ref{fig:vlsr_v_lineMass}, and the shaded region shows where $p > 0.001$. The vertical dotted line shows the expected $\sigma_v$ value of NH$_3$ for a 10 K gas with a sonic non-thermal component. \label{fig:RayTestP_vs_sigv}}
\end{figure}

We do not find a correlation between the Rayleigh test $p$-values (i.e., the randomness of the orientations) and other filament properties in Perseus, such as $\sigma_v$, $M_{\mathrm{lin}}$, and $|\nabla v_{\mathrm{LSR}}|$. As shown in Figure \ref{fig:RayTestP_vs_sigv}, however, the average randomness for filament $|\nabla v_{\mathrm{LSR}}|$ does tend to increase with $\sigma_v$ up to a value of $\sim 0.2$ km s$^{-1}$, roughly where the non-thermal component of $\sigma_v$ for a 10 K gas becomes sonic. This increase in randomness with $\sigma_v$ indicates that turbulence indeed imposes disorder to velocity fields, and hence an increase in turbulence-driven $\sigma_v$ would increase the randomness of $|\nabla v_{\mathrm{LSR}}|$ orientations.

The existence of highly non-random $|\nabla v_{\mathrm{LSR}}|$ fields in the supersonic $\sigma_{v}$ filaments, on the other hand, seems puzzling at first. While the number of these low-$p$ outliers is small, their other respective filament properties do not stand out as outliers. Moreover, given that the Rayleigh test $p$-values anti-correlate with sample sizes for a given non-random distribution, due to signal-to-noise ratio effects, the high $p$-values we measured in these filaments are likely robust and not an artifact. The sample sizes of these filaments, all being $>100$ pixels per filament, further support this claim. 

Considering all the supersonic filaments in Perseus are found exclusively at the center of NGC 1333, the high non-randomness in the $|\nabla v_{\mathrm{LSR}}|$ within this sub-population is likely a product of their environment. The fact that such randomness does not correlate with other filament properties further reinforces this claim. The local magnetic field, for example, can be stronger at this location due to the global infall (e.g., \citealt{Walsh2006}; \citealt{ChenM2020}) pinching the field, similar to what was found locally in NGC 1333 (e.g., \citealt{Doi2021}). Such an enhanced magnetic field can resist turbulence and maintain order on the velocity field. Indeed, the magnetic field and velocity gradients are well aligned towards bubble-compressed regions of NGC 1333 \citep{Dhabal2019, ChenMike2024}, even though the relationship between the two fields in NGC 1333 is messier in general, likely due to projection effects \citep{ChenMike2024}. Follow-up studies with synthetic observations of simulations will be needed to understand the effects of such projections on observations. 

It is worth noting that the supersonic filaments in NGC 1333 likely acquired a significant amount of turbulence from their environment. For example, global infall and proto-cluster feedback can inject substantial turbulence into the center of NGC 1333. Such environmental turbulence can drive these filaments' internal $\sigma_{v}$ into the supersonic regime in addition to the filament- and core-scale accretion discussed in Section \ref{subsec:cores}. While filaments may initially form as sonic entities due to the incompressibility of subsonic gases \citep{Federrath2016}, such a formation mechanism does not preclude filament-scale accretion from raising filament $\sigma_{v}$ above the sonic level alone. Nevertheless, the fact that most of these supersonic filaments have highly non-random $|\nabla v_{\mathrm{LSR}}|$ orientations does suggest they are experiencing enhanced order-imposing forces. These forces can include a stronger magnetic field resulting from a clump's global collapse, a process that also adds turbulence to filaments.

%%%%%%%%%%%%%%%%%%%%%%%%%%%%%%%%%%%%%%%%%%%%%%%%%
\section{Summary and Conclusions} \label{sec:summary}

In this paper, we fit two-component spectral models to the NH$_3$ GBT observations of the Perseus Molecular Cloud obtained by the GAS survey \citep{Friesen2017} and \cite{Pineda2010}, using the \texttt{MUFASA} software \citep{ChenM2020}. We identify velocity-coherent filaments from our emission model in PPV space using the \texttt{CRISPy} software \citep{ChenM2020} and subsequently sorted our modeled velocity slabs into velocity-coherent structures associated with each filament. We measure the internal $\nabla v_{\mathrm{LSR}}$ of these filaments on the beam-resolved scale ($\sim 0.05$ pc) and subsequently decompose them into components that are parallel and perpendicular relative to the local orientation of the filament spine. We calculate the mass enclosed within the boundaries of our velocity-coherent filaments using the {\em Herschel}-derived H$_2$ column density map of Perseus \citep{Pezzuto2021} to evaluate the gravitational stability of these filaments.

Our main results are summarized as follows:

\begin{enumerate}
  \item The median filament $M_{\mathrm{lin}}$ of each Perseus clump tightly correlates with the N(H$_2$) PDF power-law slope measured by \cite{Pezzuto2021} using the methods of \cite{Sadavoy2014}. This result indicates that filaments co-evolve with their host clump and that filament growth is driven by clumps concentrating their mass toward higher-density structures.

  \item All the Perseus filaments we identified from the NH$_3$ data have $M_{\mathrm{lin}}$ values greater than the thermally critical value of $M_{\mathrm{lin, crit}} \simeq 16$ M$_\odot$ pc$^{-1}$ for a 10 K isothermal cylinder in hydrostatic equilibrium. Most of these filaments have $M_{\mathrm{lin}} > 2 M_{\mathrm{lin, crit}}$.
  
  \item Most of the Perseus filaments have non-thermal velocity dispersions that are subsonic. Even in the supersonic cases, the median $\sigma_v$ of the filaments are all $< 0.26$ km s$^{-1}$, which is unlikely to provide significant pressure support for the filaments. Indeed, when the observed $\sigma_v$ is used to calculate the virially critical line mass, $M_{\mathrm{lin, vir}}$, all the Perseus filaments are either trans- or super-critical.
  
  \item The $\sigma_v$, $M_{\mathrm{lin}}$, $|\nabla v_{\mathrm{LSR}, \perp}|$, and $|\nabla v_{\mathrm{LSR}, \parallel}|$ values in the Perseus filaments all correlate positively with each other. The only exception is between $M_{\mathrm{lin}}$ and $|\nabla v_{\mathrm{LSR}, \parallel}|$, which indicates that $M_{\mathrm{lin}}$ is unlikely the sole driver behind all the kinematic trends we find in the Perseus filaments, despite being seemingly responsible for the radial-accretion-driven kinematics. 
  
  \item The numbers of cores and Class 0/I YSOs per unit length correlate with $\sigma_v$ and $M_{\mathrm{lin}}$. Such correlations suggest the physical processes that predetermine fragmentation spacing during the thermally subcritical phase may also govern the subsequent evolution of $M_{\mathrm{lin}}$ in the supercritical phase. Alternatively, filament fragmentation spacing may not be governed solely by the initial conditions seeded during the subcritical phase, as predicted by isolated filament models (e.g., \citealt{Inutsuka1992}). In this scenario, fragmentation space can co-evolve with $M_{\mathrm{lin}}$ well into the supercritical phase, much like those found by \cite{Chira2018} in their MHD cloud simulations.
  
  \item The $\nabla v_{\mathrm{LSR}}$ orientations, measured on the $\sim 0.05$ pc scale, are not randomly distributed in filaments. These orientations often appear to be bimodally or multimodally distributed, and their individual modes do not typically run in parallel or perpendicular to the filament spines. This result suggests that gas flows concerning filaments are not uniformly organized on the filament scale and are unlikely to be governed solely by the self-gravity of filaments.

\end{enumerate}

We conducted one of the first large sample studies of star-forming filaments' local $\nabla v_{\mathrm{LSR}}$ fields across a molecular cloud, focusing on the orientation and magnitudes of $\nabla v_{\mathrm{LSR}}$ vectors measured on the $\sim 0.05$ pc scale. While the $\nabla v_{\mathrm{LSR}}$ in the Perseus filaments are not randomly oriented, they do not show behavior naively expected of accretion flow onto and along filaments. The gas kinematics revealed by our $\nabla v_{\mathrm{LSR}}$ measurements on these small scales are inherently complex. Indeed, in-depth studies of $\nabla v_{\mathrm{LSR}}$ using the same data for Perseus NGC 1333 \citep{ChenM2020} and higher-resolution data for NGC 1333 \citep[$0.02$ pc;][]{ChenMike2024} and B5 \citep[$0.009$ pc;][]{ChenMCY2022} revealed a wealth of complex kinematic structures on smaller scales.

The increase of $\sigma_v$ and $|\nabla v_{\mathrm{LSR}, \perp}|$ with $M_{\mathrm{lin}}$ in the thermally supercritical filaments suggests these filaments are growing continuously via accretion, similar to the scenario proposed by \cite{Arzoumanian2013}. Given the complex behavior of the $\nabla v_{\mathrm{LSR}}$ fields, particularly their orientations, these accretion flows are likely highly inhomogeneous, similar to those reported in simulations by \cite{Clarke2017}. Furthermore, the correlation between the number of cores per unit length and $M_{\mathrm{lin}}$ better matches the behavior of filaments simulated within a realistic cloud environment (e.g., \citealt{Chira2018}) than those simulated in isolation (e.g., \citealt{Inutsuka1992}). These results indicate that core formation in filaments is significantly influenced by the environment and is not solely governed by perturbation growth in the early, thermally subcritical phase.

\begin{acknowledgments}
MCC, JDF, and HK acknowledge the financial support of a Discovery Grant from NSERC of Canada. MCC further acknowledges the support of CGS D from NSERC Canada. AG acknowledges support from NSF grants AST 2008101 and CAREER 2142300. AP acknowledges the support by the Russian Ministry of Science and Education via the State Assignment Contract FEUZ-2020-0038. The Green Bank Observatory is a facility of the National Science Foundation operated under a cooperative agreement by Associated Universities, Inc. The National Radio Astronomy Observatory is a facility of the National Science Foundation operated under a cooperative agreement by Associated Universities, Inc. MCC would like to further thank Pho Tru for powering many nights of paper writing, Cleo ``Babas'' for being the fluffiest sweet little editor, and Penny ``Babby'' for the nightly injections of angular momentum support.
\end{acknowledgments}


\vspace{5mm}
\facilities{GBT(KFPA), {\em Herschel}(SPIRE and PACS)}
%\facilities{HST(STIS), Swift(XRT and UVOT), AAVSO, CTIO:1.3m,CTIO:1.5m,CXO}

          
\software{astropy \citep{Robitaille2013},
    CRISPy \citep{ChenM2020},
    MUFASA \citep{ChenM2020},
    PySpeckit (\citealt{Ginsburg2011}; \citealt{Ginsburg2022})
    }
    % LicPy \citep{Rufat2018},



\appendix

\section{Kinematic Maps}
\label{appendix:kin_maps}

\begin{figure*}[t!]
\centering
\includegraphics[width=0.8\textwidth, trim={0 0 0 0}]{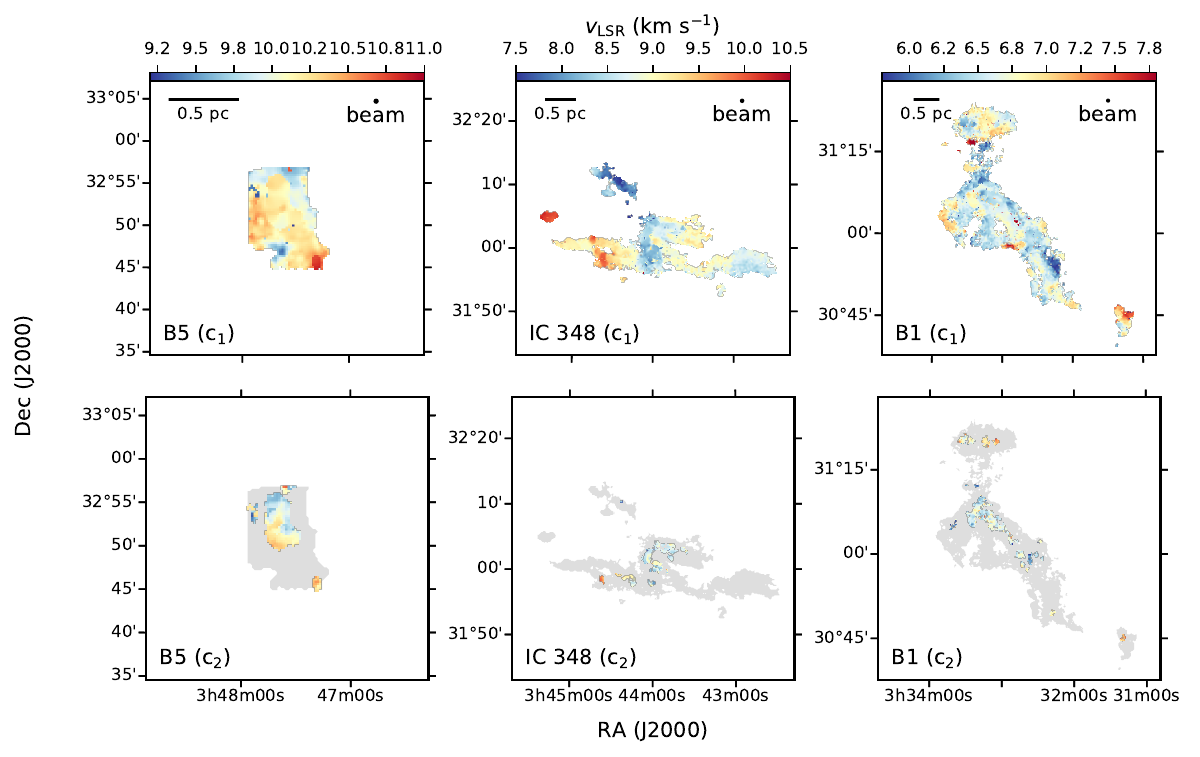}
\caption{The NH$_3$-derived v$_{\mathrm{LSR}}$ maps of Perseus B5, IC348, and B1, in units of km s$^{-1}$. The v$_{\mathrm{LSR}}$ values of the component with the narrowest $\sigma_\mathrm{v}$ (c$_1$) are shown in the first row. The v$_{\mathrm{LSR}}$ values of the widest $\sigma_\mathrm{v}$ component (c$_2$) are shown in the second row, with the grey shaded region showing where NH$_3$ are robustly detected. \label{fig:vlsr_maps_PerE}}
\end{figure*}

\begin{figure*}[t!]
\centering
\includegraphics[width=0.8\textwidth, trim={0 0 0 0}]{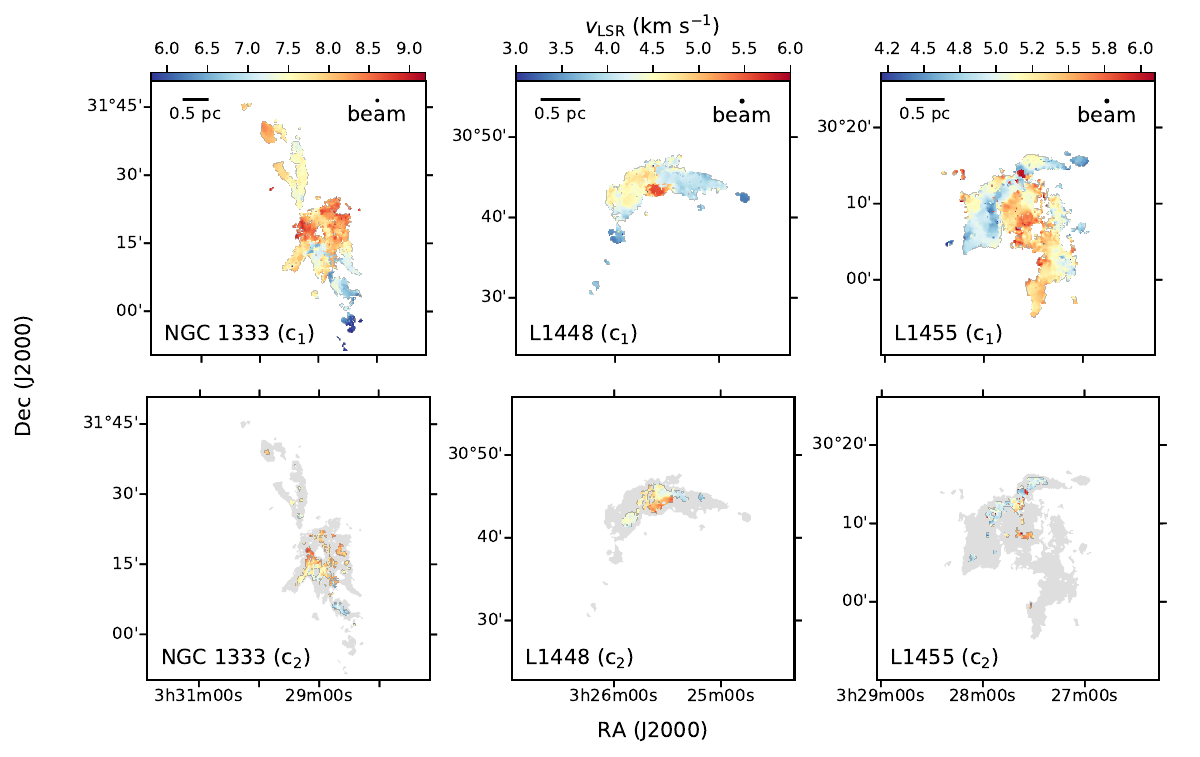}
\caption{Same as Figure \ref{fig:vlsr_maps_PerE}, but v$_{\mathrm{LSR}}$ showing maps of Perseus NGC 1333, L1448, and L1455, instead. \label{fig:vlsr_maps_PerW}}
\end{figure*}

\begin{figure*}[t!]
\centering
\includegraphics[width=0.8\textwidth, trim={0 0 0 0}]{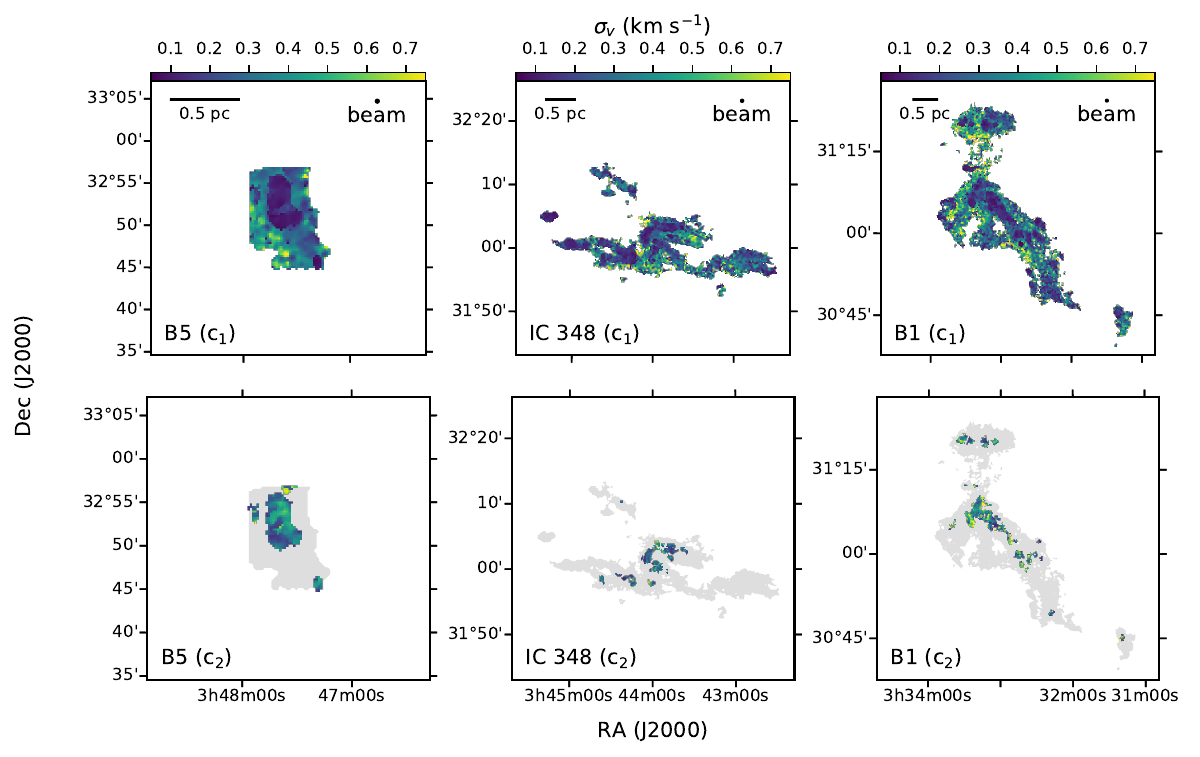}
\caption{Same as Figure \ref{fig:vlsr_maps_PerE}, but showing maps of $\sigma_\mathrm{v}$, in units of km s$^{-1}$, instead. \label{fig:sigv_maps_PerE}}
\end{figure*}

\begin{figure*}[t!]
\centering
\includegraphics[width=0.8\textwidth, trim={0 0 0 0}]{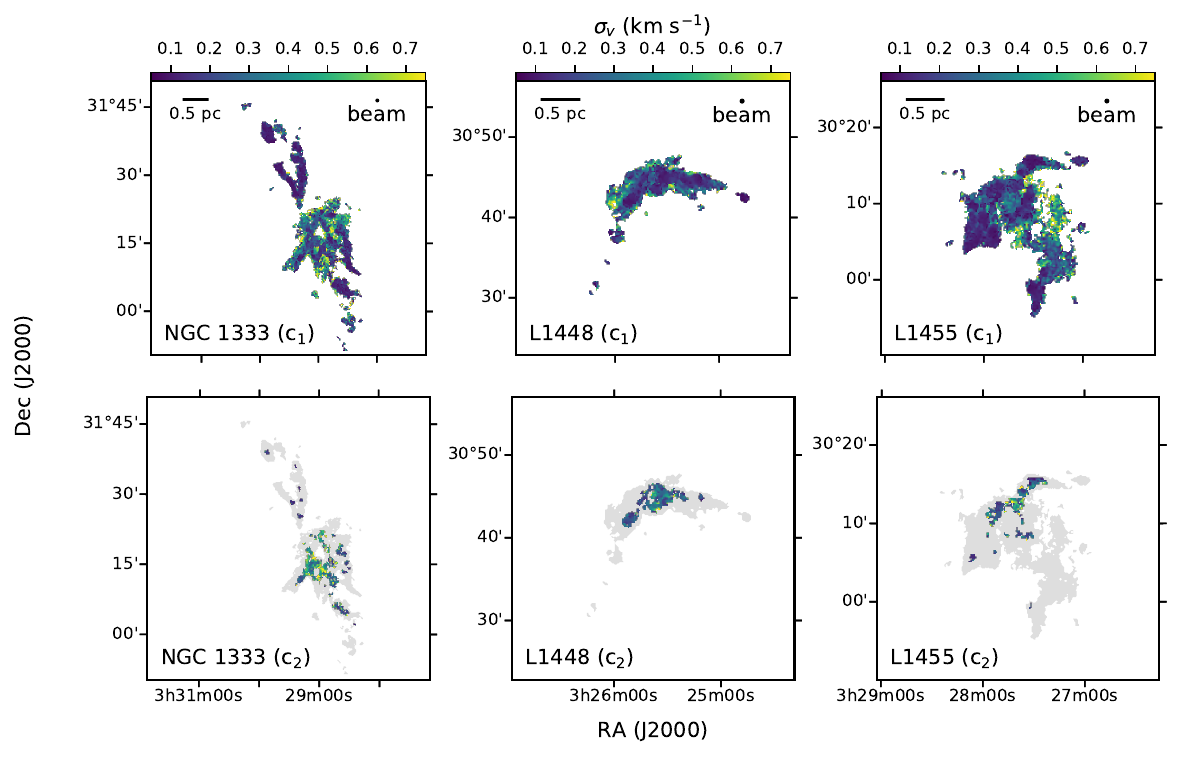}
\caption{Same as Figure \ref{fig:vlsr_maps_PerW}, but showing maps of $\sigma_\mathrm{v}$, in units of km s$^{-1}$, instead. \label{fig:sigv_maps_PerW}}
\end{figure*}

Figures \ref{fig:vlsr_maps_PerE}, \ref{fig:vlsr_maps_PerW}, \ref{fig:sigv_maps_PerE}, and \ref{fig:sigv_maps_PerW} show NH$_3$-derived v$_{\mathrm{LSR}}$ and $\sigma_{\mathrm{v}}$ maps of the six Perseus clumps. The maps are organized such that the fits-derived component with the lowest $\sigma_{\mathrm{v}}$ value in a given pixel (c$_1$) is shown in the first row and the component with the highest $\sigma_{\mathrm{v}}$ value (c$_2$) in the second row.

%=========================================================
\section{Velocity Gradient Maps}
\label{appendix:vgrad_maps}

\begin{figure*}[t!]
\centering
\includegraphics[width=0.8\textwidth, trim={0 0 0 0}]{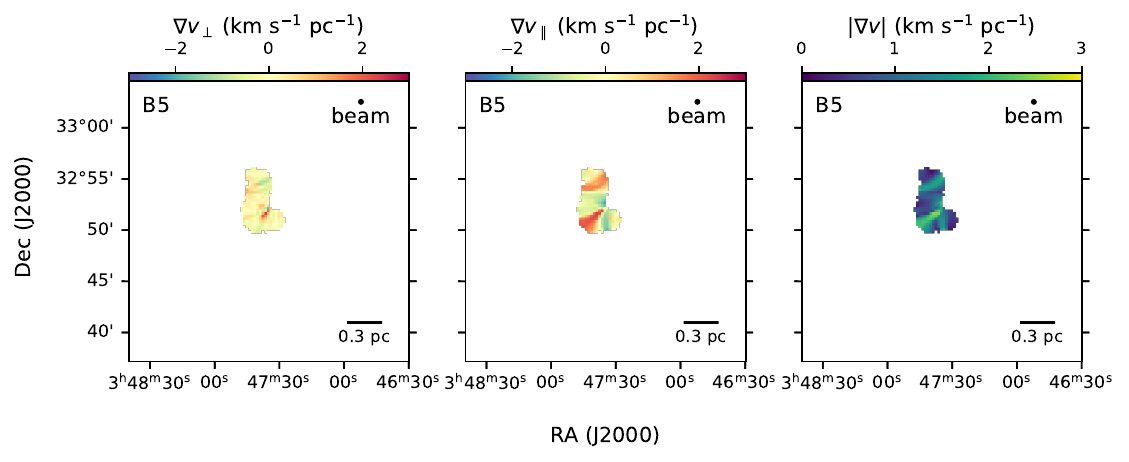}
\caption{The perpendicular component (left), the parallel component (center), and the magnitude (right) maps of $\nabla v_\mathrm{LSR}$ in the Perseus B5 filament. \label{fig:B5_vgrad_maps}}
\end{figure*}

\begin{figure*}[t!]
\centering
\includegraphics[width=0.8\textwidth, trim={0 0 0 0}]{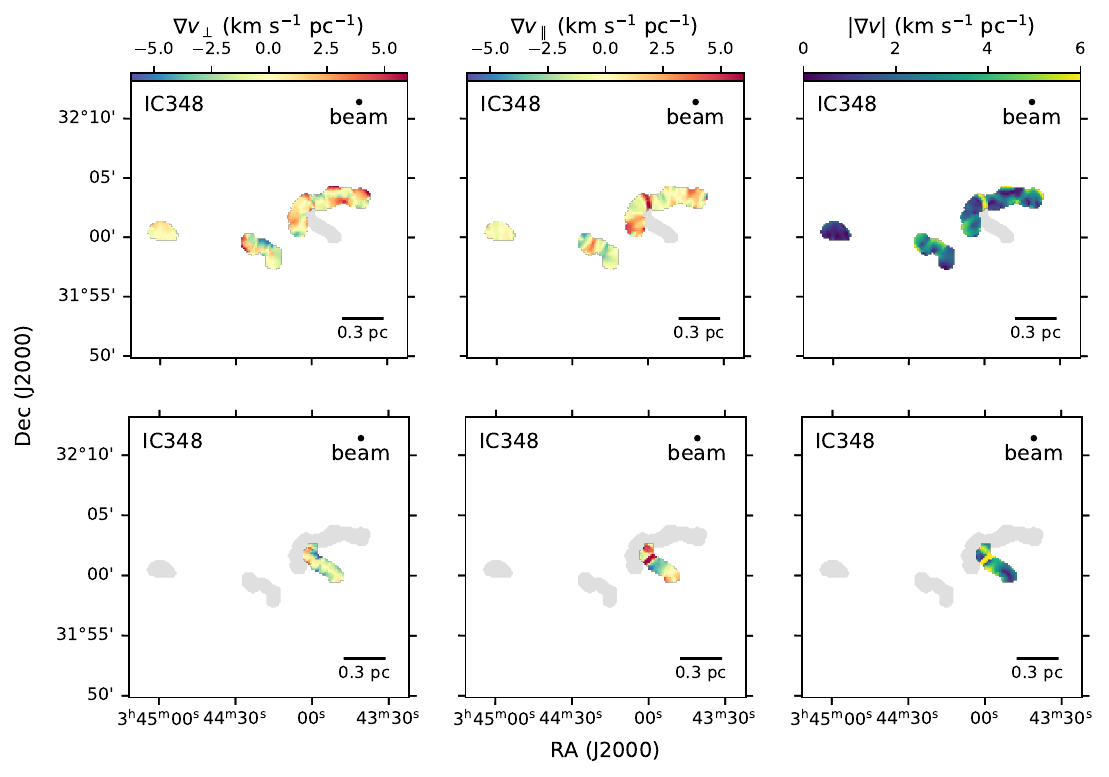}
\caption{The perpendicular component (left), the parallel component (center), and the magnitude (right) maps of $\nabla v_\mathrm{LSR}$ in the Perseus IC348 filaments. These filaments are displayed over two rows to prevent overlapping maps from obscuring one another. The footprints of the filaments not shown in each row are displayed in grey.\label{fig:IC348_vgrad_maps}}
\end{figure*}

\begin{figure*}
\centering
\includegraphics[width=0.8\textwidth, trim={0 0 0 0}]{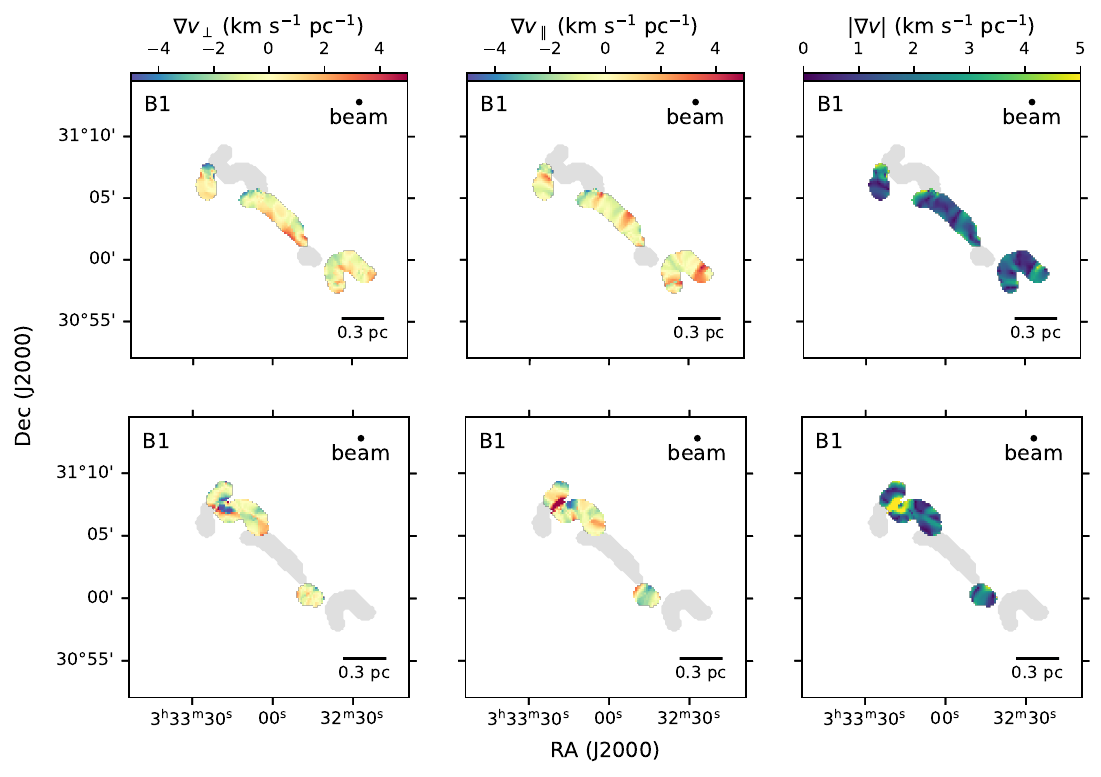}
\caption{Same as Figure \ref{fig:IC348_vgrad_maps} but for Perseus B1. \label{fig:B1_vgrad_maps}}
\end{figure*}

\begin{figure*}
\centering
\includegraphics[width=0.8\textwidth, trim={0 0 0 0}]{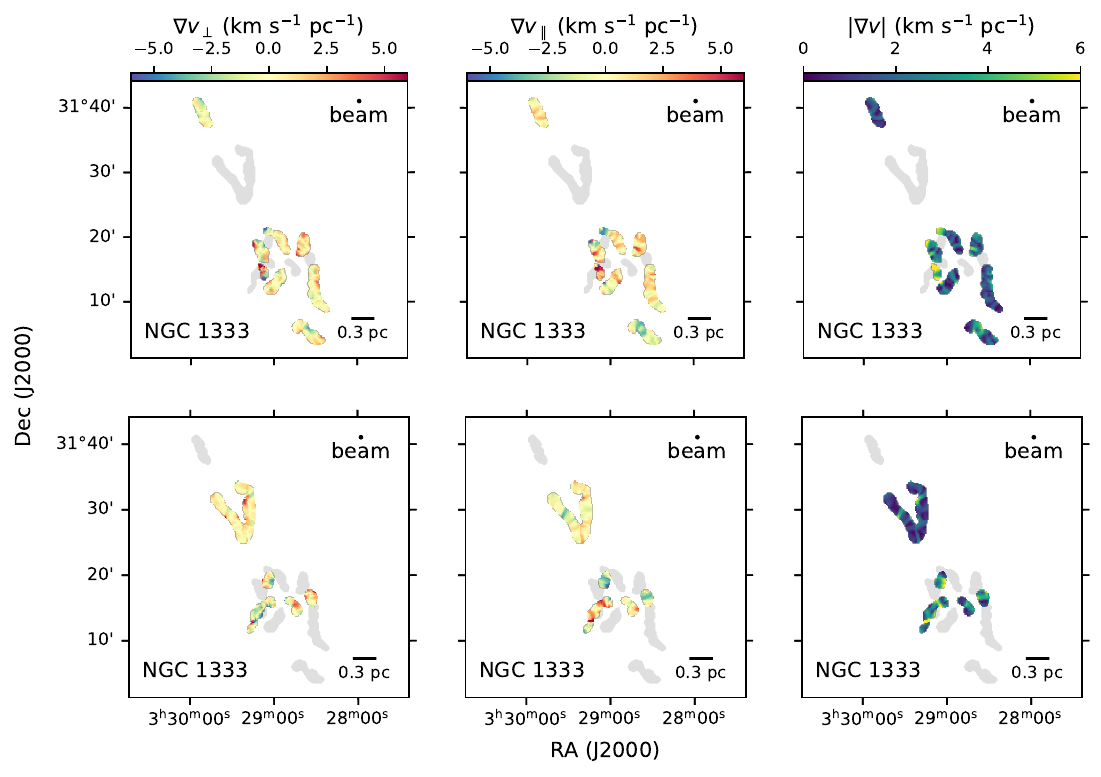}
\caption{Same as Figure \ref{fig:IC348_vgrad_maps} but for Perseus NGC 1333. \label{fig:NGC1333_vgrad_maps}}
\end{figure*}

\begin{figure*}[t!]
\centering
\includegraphics[width=0.8\textwidth, trim={0 0 0 0}]{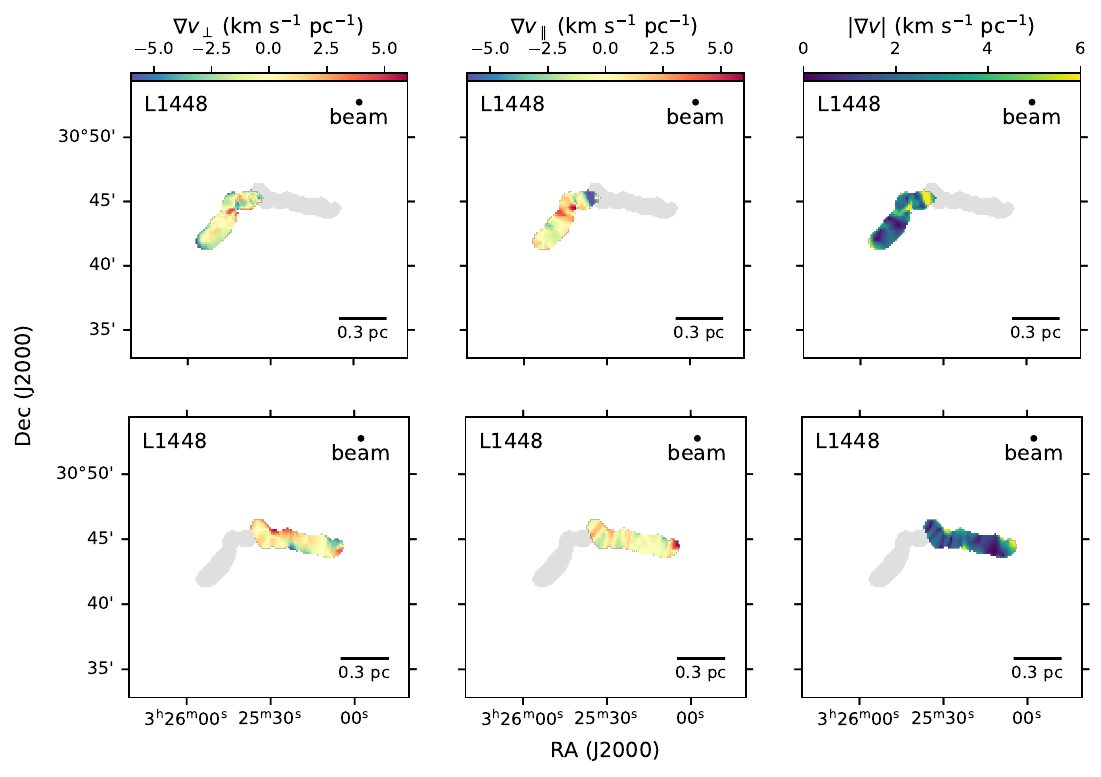}
\caption{Same as Figure \ref{fig:IC348_vgrad_maps} but for Perseus L1448. \label{fig:L1448_vgrad_maps}}
\end{figure*}

\begin{figure*}[t!]
\centering
\includegraphics[width=0.8\textwidth, trim={0 0 0 0}]{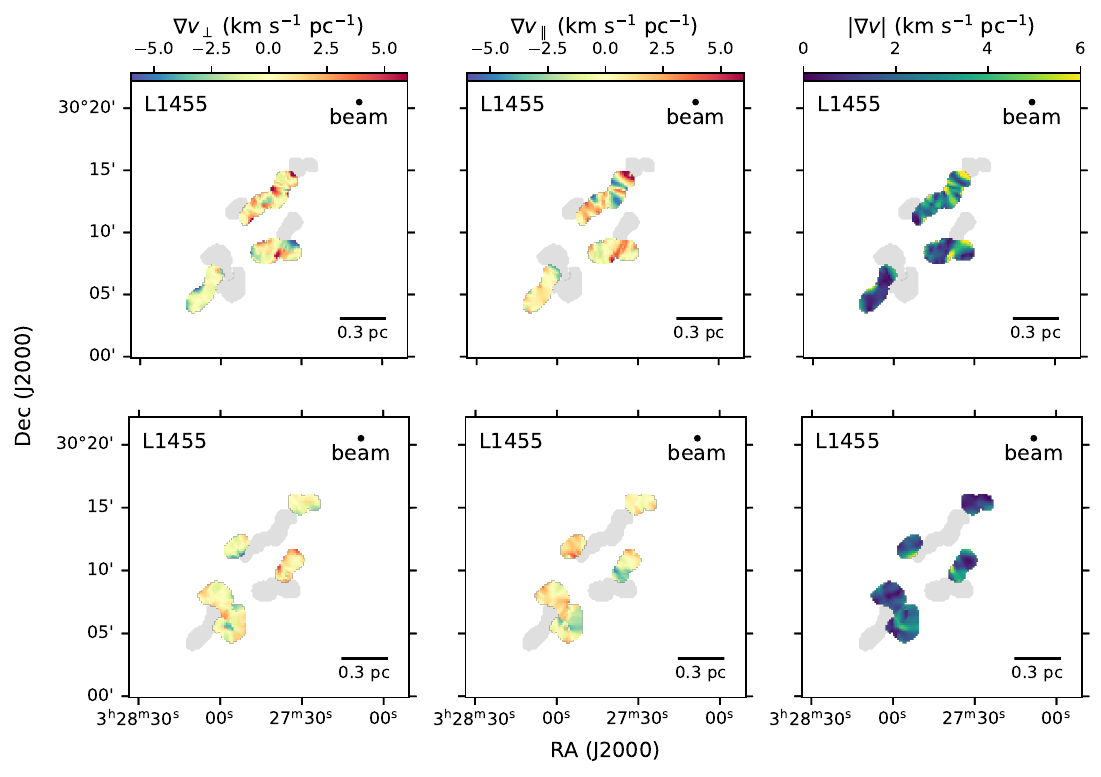}
\caption{Same as Figure \ref{fig:IC348_vgrad_maps} but for Perseus L1455. \label{fig:L1455_vgrad_maps}}
\end{figure*}

Figures \ref{fig:B5_vgrad_maps}, \ref{fig:IC348_vgrad_maps}, \ref{fig:B1_vgrad_maps}, \ref{fig:NGC1333_vgrad_maps}, \ref{fig:L1448_vgrad_maps}, and \ref{fig:L1455_vgrad_maps} show maps of $\nabla v_\mathrm{LSR}$'s perpendicular component, parallel component, and magnitude for velocity-coherent filaments we identified in each of the six Perseus clumps.

%Not needed?


\bibliography{sample631}{}
\bibliographystyle{aasjournal}



\end{document}